\documentclass[prx,superscriptaddress,twocolumn,twoside,floatfix,pra,a4paper,nofootinbib]{revtex4-2} 
\usepackage{times}
\usepackage{epsfig}
\usepackage{amsfonts}
\usepackage{amsmath}
\usepackage{array}
\usepackage{amssymb,amsthm}
\usepackage{color}
\usepackage{multirow}
\usepackage{braket}
\usepackage{bbm}
\usepackage{latexsym}
\usepackage{amsfonts}
\usepackage{mathrsfs}
\usepackage{natbib}
\usepackage{verbatim}
\usepackage{gensymb}
\usepackage{caption}
\usepackage{subcaption}
\usepackage{subcaption}
\usepackage{graphicx}
\usepackage{ dsfont }
\usepackage[colorlinks=true,linkcolor=blue,citecolor=magenta,urlcolor=blue]{hyperref}
\usepackage[qm]{qcircuit}
\allowdisplaybreaks

\usepackage{multirow}

\begin{document}

\newcommand{\qc}[1]{Circuit~(\hyperref[qc:#1]{\ref*{qc:#1}})}
\newcounter{qcnum}
\newcommand{\qcref}[1]{\refstepcounter{qcnum}\tag{\theqcnum}\label{qc:#1}}


\title{How ``Quantum'' is your Quantum Computer? \\ Macrorealism-based Benchmarking  via Mid-Circuit Parity Measurements}

\author{Ben Zindorf}
\email{ben.zindorf.19@ucl.ac.uk}
\affiliation{Department of Physics and Astronomy, University College London, Gower Street, London WC1E 6BT, England, United Kingdom}

\author{Lorenzo Braccini}
\email{lorenzo.braccini.18@ucl.ac.uk}
\affiliation{Department of Physics and Astronomy, University College London, Gower Street, London WC1E 6BT, England, United Kingdom}

\author{Debarshi Das}
\email{dasdebarshi90@gmail.com}
\affiliation{Department of Physics, Shiv Nadar Institution of Eminence, Gautam Buddha Nagar, Uttar Pradesh 201314, India}
\affiliation{Department of Physics and Astronomy, University College London, Gower Street, London WC1E 6BT, England, United Kingdom}

\author{Sougato Bose}
\affiliation{Department of Physics and Astronomy, University College London, Gower Street, London WC1E 6BT, England, United Kingdom}
	
\begin{abstract}

To perform meaningful computations, Quantum Computers (QCs) must scale to macroscopic levels -- i.e., to a large number of qubits -- an objective pursued by most 
quantum companies. How to efficiently test their quantumness at these scales? We show that the violation of Macrorealism (MR), being the fact that classical systems possess definite properties that can be measured without disturbances, provide a fruitful avenue to this aim. The No Disturbance Condition (NDC) -- the equality used here to test MR -- can be violated by two consecutive parity measurements on $N$ qubits and found to be independent of $N$ under ideal conditions. However, realistic noisy QCs show a quantum-to-classical transition as $N$ increases, giving a foundationally-motivated scalable benchmarking metric. 
Two methods are formulated to implement this metric: one that involves a mid-circuit measurement, probing the irreversible collapse of the wavefunction, in contrast to the reversible entanglement generated in the other.
Both methods are designed to be \textit{clumsiness-loophole free}: the unwanted classical disturbances are negligible within statistical error. Violation of MR is detected on a IBM QC up to $N = 38$ qubits, increasing $N$ by one order of magnitude over best known results of MR. Two QCs are benchmarked using the proposed NDC metric, showing a three-fold improvement in their quantumness from one generation to the next.

\end{abstract}
\maketitle



\section{Introduction \label{sec:intro}}

The essence of quantum mechanics lies in quantum interference and its disappearance upon intermediate measurement, revealing the wave-particle duality of quantum systems~\cite{feynman_lecture_3}. These genuine quantum features, which cannot be explained using classical mechanics, are direct consequences of the quantum mechanical postulates: the superposition principle and the measurement postulate. Recognised as resources for computation~\cite{Benioff_1980, Feynman_1982}, superposition and entanglement have enabled a range of quantum algorithms that outperform their classical counterparts~\cite{Shor_1994, Grover_1997,harrow_quantum_2009,nielsen_quantum_2010,Schmidhuber_Quartic}. Similarly, the non-unitary discontinuous change of quantum states under measurement is at the root of quantum error correction~\cite{knill_theory_1997,gottesman_stabilizer_1997,chiaverini_realization_2004,singh_mid-circuit_2023,google_quantum_ai_2023}, state teleportation~\cite{teleportation_bennett_1993,bouwmeester_etal_1997,gottesman_chuang_1999,kimble_quantum_2008,measurement_baumer_2025}, and measurement-based computation and state preparation~\cite{raussendorf_one-way_2001,Raussendorf_measurement_2003,Iqbal_Topological_2024}. A promising pathway to achieve these applications is via appropriate encoding and the measurement of parity of a register of physical qubits, corresponding to a logical qubit~\cite{bultink_protecting_2020,smith_minimally_2025,fellner_universal_2022,saira_entanglement_2014}.

Reliably benchmarking a large register of qubits is essential for full-scale quantum computation~\cite{lall_review_2025,wack_quality_2021}. This has motivated scalable benchmarking protocols for Quantum Computer (QC) such as quantum volume~\cite{benchmarking_volume,benchmarking_volume_2}, randomised circuit mirroring~\cite{magesan_scalable_2011,kimmel_robust_2014,boixo_characterizing_2018,cross_validating_2019}, and specific algorithmic performances~\cite{tomesh_supermarq_2022, benchmarking_variation,benchmarking_algo,Ferracin2018reducing,Ferracin2019accrediting}. However, only a few benchmarking suites investigate errors induced by mid-circuit measurements~\cite{govia_randomized_2023,hothem_measuring_2024} and, to the best of our knowledge, none of these focusses on parity-measurements. A well-motivated benchmarking metric capable of differentiating ``classical" and ``quantum" collective measurements is still lacking. In this work, we present a scalable protocol, the first of its kind, to quantitatively certify the \textit{quantumness}, i.e. nonclassicality, of a QC as a collective macroscopic entity, leveraging the quantum superposition principle alongside quantum measurement-induced disturbance, with the main focus on mid-circuit parity measurements. 

In the field of quantum foundations, philosophical axioms of theories can be used to formulate no-go theorems constraining correlations via inequalities: 
for instance, in the setting of Bell tests \cite{Bell}, correlations of entangled states cannot be reproduced by simultaneously local and realist models. Similarly, the notion of macrorealism~(MR) investigates whether a single system can be described by a realist model in which all physical observables have pre-defined values that are, in principle, unaffected by measurements~\cite{lgi1,leggett02,lgi2}. These assumptions lead to a set of necessary conditions for MR, e.g., the Leggett-Garg inequalities~\cite{lgi1,qlgi1} and the No-Disturbance Condition (NDC)\footnote{also known as the No Signaling in Time Condition.}~\cite{nsit1,nsit2,nsit3,gravity_measurement}, which constrain the time-separated correlations between outcomes of successive measurements performed on a system. Quantum systems violate the MR assumptions via the superposition principles and measurement postulate. On the one hand, violation of MR has been experimentally certified for a single qubit~\cite{Athalye_2011,knee2012violation, nsit1}, and up to five qubits with QCs \cite{LGIQC1,LGIQC2,LGIQC3,LGIQC4,LGIQC5}, without approaching a macroscopic (scalable) limit. On the other hand, analyses that show the quantum-to-classical transition via MR as the quantum system becomes macroscopic have been purely theoretical~\cite{large_spin1,schlosshauer_quantum_2007,MassMR2024,classicality_emergence}. 


Here we treat a QC as a single macroscopic quantum system, composed of many fundamental elements, to experimentally detect the violation of MR through NDC, ruling out any classical description of the QC. The NDC protocol is realised by consecutive parity measurements of $N$ qubits, such that, in the ideal case, the violation can be made independent from $N$ and thus applicable to the macroscopic limit of a QC. This metric probes the quantumness of a QC as a whole, measuring the collective coherence of the computer concomitantly with the discontinuous collapse of its wave function, involving mid-circuit measurement. Thus, our metric investigates all postulates of quantum mechanics. Two specific methods have been employed to eliminate the \textit{clumsiness loophole}, which is the measurement-induced disturbances of classical systems introduced by experimental noise. As the number of qubits ($N$) increases, QCs exhibit a quantum-to-classical transition that is found to be a well-motivated scalable benchmark based on the foundational notions of quantumness: \textit{At which number of qubits $N_{\text{NDC}}$ a QC stops behaving as a quantum system and becomes a macrorealistic classical one?} Violation of MR is detected up to $38$ qubits with \texttt{ibm\_marrakesh} QC, and its transition to classicality is compared to the one of \texttt{ibm\_brisbane} (up to $14$ qubits), a QC of the previous generation.

\section{The Parity NDC Protocol \label{sec:ideal_case}}

Let us start by formally introducing the assumptions of MR, given by: (1) \textit{Realism per se:} At any instant, even if unobserved, a system is definitely in one of its possible states with all its observable properties having definite values. (2) \textit{Noninvasive measurability:} It is possible, at least in principle, to determine which of the states the system is in by ensuring that the measurement-induced disturbance is arbitrarily small, thus not affecting the state or the subsequent time evolution of the system. The two-time NDC -- the condition chosen to test MR -- involves two sequential measurements of a dichotomic observable (with $\pm$ outcomes) on an evolving system at different instants, $t_1$ and $t_2$, where $t_1 < t_2$. Mathematically, the NDC can be expressed as
\begin{align}
	V_{\pm}  =	 P_2(\pm) - \left[  P_{1,2}(+,  \pm) +  P_{1,2}(-,\pm) \right]  = 0.
	\label{eq:ndc}
\end{align}
where, for example, $P_{2}(+)$ is the probability of the outcome $+$ at the instant $t_2$, without any measurement at $t_1$, and $P_{1,2}(+, -)$  is the joint probability of obtaining the outcomes $+$ at instant $t_1$ and $-$ at instant $t_2$. In the following, we will consider $V := V_+ = - V_-$, where the second equality follows from the normalization of probabilities. 

This NDC stipulates that the probability of obtaining a specific measurement outcome at a later time ($t_2$) must be independent of whether any measurement was performed at an earlier time ($t_1$). This condition should hold for any classical system, where any unwanted measurement-induced disturbances can, in principle, be made arbitrarily small (see Sec.~\ref{sec:implementation}). When probing a quantum system, the violation $V$ quantifies the quantum measurement-induced collapse of the wavefunction, together with its coherence, testing inherent and fundamental quantum properties. The parity NDC protocol -- that this work proposes -- implements and measures the NDC violation of an ensemble of $N$ qubits via parity measurements on a QC. 
In the following, we present the main mathematical results using the quantum computing language (for detailed proofs, we refer to our parallel work \cite{LargeSpinParallel}, where a qubit ensemble interacts through a single monitored resonator and noise analysis is performed). 

Let us consider $N$ qubits with Hilbert space $\mathcal{H} \sim \left( \mathds{C}^{2} \right)^{\otimes N}$ with the complete computational basis defined as the set of orthonormal states $\ket{J} = \bigotimes_{i = 1}^{N} \ket{j_i}$, with $j_i = \{0, 1\}$ and $J$ being the string $\{j_1, ..., j_N\}$, labeling a state. The completeness of the basis can be expressed as $\sum_{J \in \mathcal{S}} \ket{J} \bra{J} = \mathds{1}$, where $\mathcal{S}$ is the set of all possible strings.   The parity measurement projects any state to the even or odd subspaces of the $N$-qubits Hilbert space. Formally, the even subspace is spanned by the vectors $\{ \ket{J} \}$ with $J \in \mathcal{S}_{\text{e}}$, where $\mathcal{S}_{\text{e}}$ is the set of strings such that $\sum_{i = 1}^N j_i$ is even (and similarly for the odd case, with $\mathcal{S}_{\text{o}}$). Then, such a measurement is described by the two projectors
\begin{equation}
\label{eq:POVM}
    \Pi_{\text{e}} = \sum_{J \in \mathcal{S}_{\text{e}}} \ket{J} \bra{J}\;,  \hspace{1cm} 
    \Pi_{\text{o}} = \sum_{J \in \mathcal{S}_{\text{o}}} \ket{J} \bra{J} \;.
\end{equation}
One can note that it is indeed a Positive Operator-Valued Measure (POVM) as $\Pi_{\text{e}} + \Pi_{\text{o}} = \mathds{1}$, from the fact that $\mathcal{S}_\text{e} \cup \mathcal{S}_\text{o} = \mathcal{S}$ and the completeness of the computational basis. 

\begin{figure}[t!]
    \centering
\includegraphics[width=0.45\textwidth]{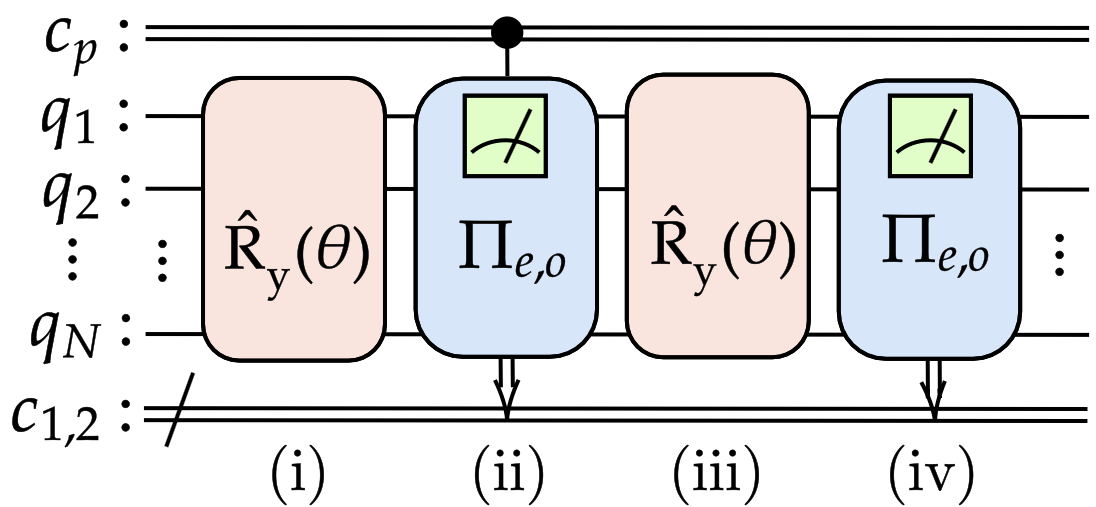}
    \caption{Diagram of the circuit implementing the parity NDC protocol on $N$-qubits via (i) an initial rotation, (ii) a parity measurement classically-controlled by bit $c_p$, (iii) a second rotation, and (iv) a final parity measurement.\label{fig:Control_Parity}}
\end{figure}

The parity NDC protocol can be achieved through four stages outlined below and represented in Fig. \ref{fig:Control_Parity}.

(i) \textit{State Preparation}: Each qubit is initialised in the state $\ket{0}$ and undergoes an identical rotation by angle $\theta$ around the $y$-axis. The initial $N$-qubit state $\ket{0}^{\otimes N} $ evolves to $ \ket{\theta}_N  = \hat{\mathcal{R}}_y  (\theta)\ket{0}^{\otimes N}$, where $\hat{\mathcal{R}}_y (\theta)$ represents the $y$ rotation of all $N$ qubits. This creates a superposition in the $z$-base given by
\begin{equation}
\label{eq:rotates_state}
    \ket{\theta}_N =   \frac{1}{2^{N/2}} \left( \cos\left(\theta/2\right) \ket{0} + \sin{\left(\theta/2\right)} \ket{1} \right)^{\otimes N} \;.
\end{equation}

(ii) \textit{Classically-Controlled Parity Measurement}: To test the NDC, two sub-protocols must be carried out: the single-measurement and a double-measurement parts used to determine the probabilities $P_2$ and $P_{1,2}$ of Eq.~(\ref{eq:ndc}), respectively. In the former, only a final parity measurement is performed, while the latter includes an intermediate parity measurement. Abstractly, the choice of whether or not to perform the intermediate measurement can be viewed as a parity measurement controlled by a classical bit, $c_p$ taking the values $0$ or $1$. 

On the one hand, if the classical bit is $1$, the parity measurement is performed on the $N$ qubits and the two probabilities of outcomes $+$ and $-$ (i.e., even and odd respectively) are 
\begin{equation}
\label{eq:prob_1_ideal}
     P_1(\pm) = \frac{1}{2} \pm \frac{1}{2}  (\cos \theta)^N \; .
\end{equation}
Due to the measurement, the wavefunction of the QC collapses to the unnormalized post-measurement states: $\ket{\psi}^{\text{post}}_+ =  \Pi_e \ket{\theta} := \ket{\theta^{ \text{e} } }$ and $\ket{\psi}^{\text{post}}_- = \Pi_o \ket{\theta} := \ket{\theta^{ \text{o} } }$, with $\ket{\theta^{ \text{e} } }$ and $\ket{\theta^{ \text{o} } }$ being often named even and odd spin-coherent states of the $N$ qubits respectively~\cite{LargeSpinParallel}. The squared norms of these states are equal to the probabilities of Eq.~(\ref{eq:prob_1_ideal}). On the other hand, if the classical bit is equal to $0$, the parity measurement is not performed, and this step is trivial (leaving the qubits in the same quantum state of Eq.~\ref{eq:rotates_state}).

(iii) \textit{Second Rotation}: An identical rotation of angle $\theta$ is performed on all qubits. If the intermediate measurement was performed, the post-measurements states are rotated to the state $\hat{\mathcal{R}}_y (\theta)  \ket{\psi}^{\text{post}}_\pm $. If the intermediate measurement was not performed, the rotated state is given by Eq.(\ref{eq:rotates_state}) with $\theta$ being replaced by $2 \theta$, as the two rotations are around the same axis.

(iv) \textit{Final Measurement}: For both parts of the protocol, the system undergoes a final parity measurement on all the qubits.

In the ideal case, by comparing the single-measurement and the double-measurement protocols presented above, the violation of NDC given by Eq.~(\ref{eq:ndc}) is found to be
\begin{equation}
\label{eq:violation_ideal}
    V = \frac{1}{4} \left[1- (\cos 2 \theta )^N  \right] \;. 
\end{equation}
For $\theta = \pi/4$, the violation of NDC is independent of the number of qubits, with $V = 1/4 \; \forall N$. This implies that it is possible, in principle, to detect quantumness of an ensemble of qubits, independent of its size, thus probing quantum mechanics in the macroscopic limit as the total number of qubits increases. This parameter choice is the one that will be used in the following.  
 In Fig. \ref{fig:angles}, the ideal violation as function of the angle $\theta$ is reported (along with the one detected on IBM QCs described later).

\begin{figure*}[t!]
    \centering
\includegraphics[width=0.97\textwidth]{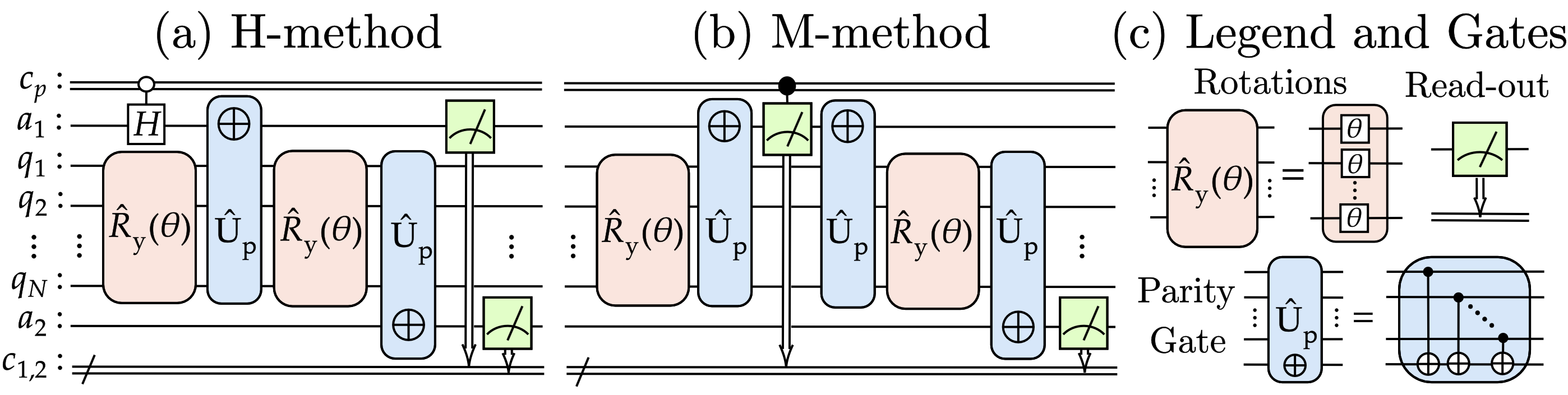}
    \caption{Circuits implementing (a) the H-method and (b) the M-method of the parity NDC protocol on an ensemble of $N$ qubits ($q_{1},...,q_N$) with two ancilla qubits ($a_{1}$, $a_2$) for implementing $N$-qubits parity measurements with outcomes stored in $2$ classical bits ($c_{1}$ and $c_{2}$), and a classical bit ($c_p$), which controls the intermediate measurement via: (a) classically controlling a Hadamard gate applied on the ancilla $a_1$ (the gate is implemented when $c_p=0$), or (b) classically controlling a mid-circuit $Z$-measurement on the ancilla $a_1$ (the measurement is implemented when $c_p=1$). The legend is given in (c).\label{fig:circuit_1}}
\end{figure*}
 
\section{Implementations on Quantum Computers\label{sec:implementation}}

The underlying idea of this work  is to perform each parity measurement using a dedicated ancilla qubit -- two different ancilla qubits in total for the two parity measurements. Each ancilla qubit stores the parity of the $N$ qubits and then is measured in the computational basis, thereby implementing the corresponding parity measurement on the $N$ qubits. 
In order to achieve this, the ancilla is prepared in the state $\ket{0}_a$ and is entangled with the $N$-qubits through a parity gate 
$\hat{U}_{\text{p}} = \mathds{1}_a \otimes \Pi_e   +  X_a \otimes \Pi_o $, where $\Pi_i$ are given in Eq.~(\ref{eq:POVM}) and the subscript $a$ refers  to the ancilla Hilbert space. If the state of the $N$ qubits before the first parity measurement is in the superposition $\ket{\theta}_N$ of Eq.~(\ref{eq:rotates_state}), the ancila-qubits joint state after the parity entangling dynamics evolves to the cat state 
\begin{equation}
\label{eq:Parity_gate_6}
    \ket{\psi}^{\text{pre}} = \hat{U}_{\text{p}} \ket{0}_a \ket{\theta}_N = \ket{0}_a \ket{\theta^{ \text{e} } }_N + \ket{1}_a  \ket{\theta^{\text{o}} }_N \; , 
\end{equation}
where $\ket{\theta^\text{i}}_N = \Pi_i \ket{\theta}_N$, with $i \in \{e,o \}$ as defined earlier.  This can be implemented as a series of CNOT gates controlled by each qubit within the $N$ ones and with the ancilla as a target (Fig.~\ref{fig:circuit_1}). Each CNOT gate flips the ancilla if the control qubit is in state $\ket{1}$, and, thus, being applied to all $N$ qubits, the collective parity of the $N$ qubits becomes entangled with the ancilla qubit. The read-out of the parity can be done by performing a $Z$-measurement on the ancilla qubit -- the outcomes $\pm 1$ (e.g. $\ket{0}_a \leftrightarrow 1$) have probabilities given by Eq.~(\ref{eq:prob_1_ideal}) in the ideal case, and the state of the $N$ qubits collapses to $\ket{\theta^\text{e}}_N$ (even subspace) or $\ket{\theta^\text{o}}_N$ (odd subspace).  Thus, the action of $\hat{U}_p$ and the read-out process effectively implements a parity measurement of the $N$ qubits.  Stage (iv)  of the protocol  -- when the parity measurement is always performed -- is implemented accordingly, via the ancilla qubit $a_2$ which stores the measurement outcome in the classical bit $c_2$. 


The classically controlled parity measurement -- necessary for step (ii) and controlled by the classical bit $c_p$ -- can be implemented as follows. When the measurement is performed, it is applied as above, via the ancilla qubit $a_1$ and the outcome-storing bit $c_1$. However, in order to \textit{not} perform the measurement, one can proceed by canceling either (a) the effect of the entangling dynamics between the ancilla and the ensemble, or (b) the mid-circuit measurement of the ancilla. The two methods, named ``H-method" and ``M-method", respectively, investigate different aspects of a QC via two implementations of classically controlled parity measurements. Specifically, the former probes the quantumness via the entangling part of quantum measurements, while the latter includes a mid-circuit measurement and directly tests irreversibility due to the read-out process. 

However, in realistic experimental scenarios, measurement-induced disturbances can also occur for classical systems, while, in principle, it is always possible to ensure that such unwanted disturbances -- named classical disturbance (CD) -- are arbitrarily small, since no foundational principle of classical physics prohibits this. In the context of NDC, CDs take the form of a \textit{clumsiness loophole}, for which part of the violation of NDC can be attributed to CDs~\cite{Joarder_Loophole_2022,Wilde_Clumsiness_2012,Knee_strict_2016,Lambert_Experimental_2020}. The CDs must be negligible to ensure that the observed NDC violation is primarily due to the quantum measurement-induced wavefunction collapse.  To achieve a clumsiness-loophole-free implementation, the two sub-protocols (with and without the intermediate parity measurement) must be as \textit{similar} as possible to each other, i.e., the classically controlled parity measurements must have little differences between performing and not performing the measurement\footnote{Because CDs arise from the difference in noise between the single and double measurement sub-protocols of the experiment (Eq.~\ref{eq:ndc}).}. For instance, the naive choice of not performing a whole intermediate parity measurement, as previously described, introduces high CDs -- measured to amount up to $\sim 30\%$ of the total ideal violation on IBM QCs (see Fig. \ref{fig:IBM_results}) -- as the two sub-protocols have very different circuit structures, running time, and number of gates (Appendix~\ref{app:Classical_Disturbance}). Hence, we design the classically-controlled parity measurement such that the difference between the intermediately measured and unmeasured sub-protocols is only one single-qubit gate (H-method) or a single-qubit mid-circuit measurement (M-method).

In detail, the H-method involves rotating each of the $N$ qubits by an angle $\theta$, followed by a collective parity measurement, then applying another rotation of angle $\theta$ to each qubit, followed by a second collective parity measurement (implemented as previously discussed). A Hadamard gate classically controlled by $c_p$ is implemented on the ancilla ($a_1$), as shown in Fig.~\ref{fig:circuit_1}. On the one hand, when $c_p = 1$, the Hadamard is not performed and the circuit implements the double measurement sub-protocol on the ensemble. On the other hand, when $c_p = 0$, the Hadamard gate is executed on $a_1$ in parallel to the other rotations on the $N$ qubits, so that the state of the ancilla evolves from $\ket{0}_{a_1}$ to $\ket{+}_{a_1} = (\ket{0}_{a_1} + \ket{1}_{a_1})/\sqrt{2}$. In this case, the first parity measurement is not performed as the series of CNOT gates does not entangle the ancilla with the rest of the QC (as $\ket{+}_{a_1} \ket{\theta}_N$ is an eigenstate of $\hat{U}_p$), thus implementing the single measurement sub-protocol. 
The read-out measurements of both ancilla qubits ($a_{1,2}$) can be performed in parallel at the end of the circuit, as the state of the first ancilla does not change during the second rotation and the second entangling dynamics. Hence, the two sub-protocols are implemented with the same circuit up to a single-qubit Hadamard gate (This method is referred to as the H-method due to this reason), which can be performed in parallel to the other rotations of the $N$ qubits, ensuring approximately the same run times and circuit depth such that the CDs are negligible (see Fig. \ref{fig:IBM_results}). 

In the M-method, after the first rotation, the parity entangling dynamics between the $N$ qubits and the ancilla $a_1$ is always applied, followed by a classically controlled mid-circuit measurement, i.e., a mid-circuit $Z$-measurement conditional on the classical bit $c_p$. In order to cancel the effect of the entangling dynamics, the same parity entangling dynamics has to be applied again, as shown in Fig. \ref{fig:circuit_1}. If  $c_p = 1$, the read-out measurement is performed, collapsing the state of the $N$-qubits to the even and odd subspace. In this case, the second $\hat{U}_p$  ``resets" the ancilla back to its initial state ($\ket{0}_{a_1}$), while it acts trivially on the $N$ qubits, i.e. the post-measurement states evolve as $\hat{U}_p \ket{0}_{a_1}  \ket{\theta^{ \text{e} } }_N = \ket{0}_{a_1} \ket{\theta^{ \text{e} } }_N $ or $\hat{U}_p \ket{1}_{a_1}  \ket{\theta^{ \text{o} } }_N = \ket{0}_{a_1} \ket{\theta^{ \text{o} } }_N $, depending on the $\pm 1$ outcome. If the read-out measurement is not performed  ($c_p = 0$), the total dynamics becomes trivial, as $\hat{U}_p \hat{U}_p = \mathds{1}$. The end of the protocol consists of the second rotation and the final parity measurement. 
Hence, the two sub-protocols in this method are implemented with the same circuit up to a mid-circuit $Z$-measurement (it is referred to as the M-method due to this reason).

Additional care has to be taken in the implementation of the classically controlled mid-circuit measurement in the M-method, as measurements in QCs are slower processes compared to gates. In order to further decrease CDs, the classical bit $c_p$ controls on \textit{which} qubit the mid-circuit $Z$-measurement is performed: when $c_p=1$, the measurement is performed on the ancilla qubit entangled with the ensamble (thus implementing the intermediate parity measurement), and when $c_p=0$, the measurement is performed on some other uncorrelated qubit (thus not implementing the intermediate parity measurement). This ensures that the running times and circuit depth of the single measurement and the double measurement sub-protocols are the same, as the mid-circuit $Z$-measurement is always performed on some qubit. This further decreases the CDs for the M-method (see Fig.\ref{fig:IBM_results}).

The circuit presented in Fig. \ref{fig:circuit_1} includes long-range CNOT gates that can cause significant noise, spoiling the desired detection of quatumness and, in some technologies, it may not even be allowed. For this reason, equivalent circuits can be found (for the M-method and the H-method see Appendix~\ref{app:quantum_circ} and \ref{app:mid_circuit}, respectively), which implement the protocols efficiently under Linear Nearest Neighbor (LNN) connectivity: a connectivity which is more restrictive than the heavy-hex (allowed in IBM's QC) and can be directly mapped to many connectivity types. To further limit the noise, the depth of the circuit and the number of gates are decreased using circuit optimisation techniques. A lower depth -- measured in terms of the number of time slots during which several gates might be applied in parallel -- limits the overall running time and hence the effect of decoherence. A lower number of gates reduces the possible experimental operational errors. Of particular interest is the number of LNN CNOT gates, which, up to local rotations, are the entangling gates applied between nearest-neighboring qubits, as they represent the most challenging gates to perform. The most efficient implementation of the H-method found has depth $N + \mathcal{O} (1) $ and number of LNN CNOT gates is $3N + \mathcal{O} (1)$.\footnote{As a comparison, the depth and number of LNN  CNOT gates required to perform a single long-range CNOT gate over $N$ qubits is $N+O(1)$ and $4N+O(1)$, respectively, under the restriction of a constant number of ancillas \cite{kutin2007computation}.} The three entangling dynamics of the M-method increase the depth of the circuit to $1.5 N + \mathcal{O} (1) $, while the number of LNN CNOT gates is the same as that of the H-method. The running time (and operational noise) of the M-method is increased further compared to the H-method as it also includes a mid-circuit measurement.

Finally, it should be noted that, in the case of IBM's QCs, $y$-axis rotations cannot be implemented as a single gate and are decomposed into $\sqrt{X}$ gates and rotations around the $z$-axis. Thus, the gates used in the presented methods after compilation -- including the entangling gate for the parity measurements -- are a universal set.\footnote{For future benchmarking in different platforms, the axis of rotation, which can be any on the $xy$-plane, should be chosen such that the NDC protocol uses all single gates necessary for universality.} Furthermore, the specific choice of $\theta = \pi/4$ requires a single $T$ ($:=\sqrt[4]{Z}$) gate per rotation per qubit (i.e., a total of $2N$ $T$ gates), thus making the proposed methods efficient for scalable fault-tolerant QC, for which $T$ gates are the most expensive resource.

\section{Benchmarking \&  Results\label{sec:results}}

\begin{figure}[t!]
    \centering
\includegraphics[width=0.48\textwidth]{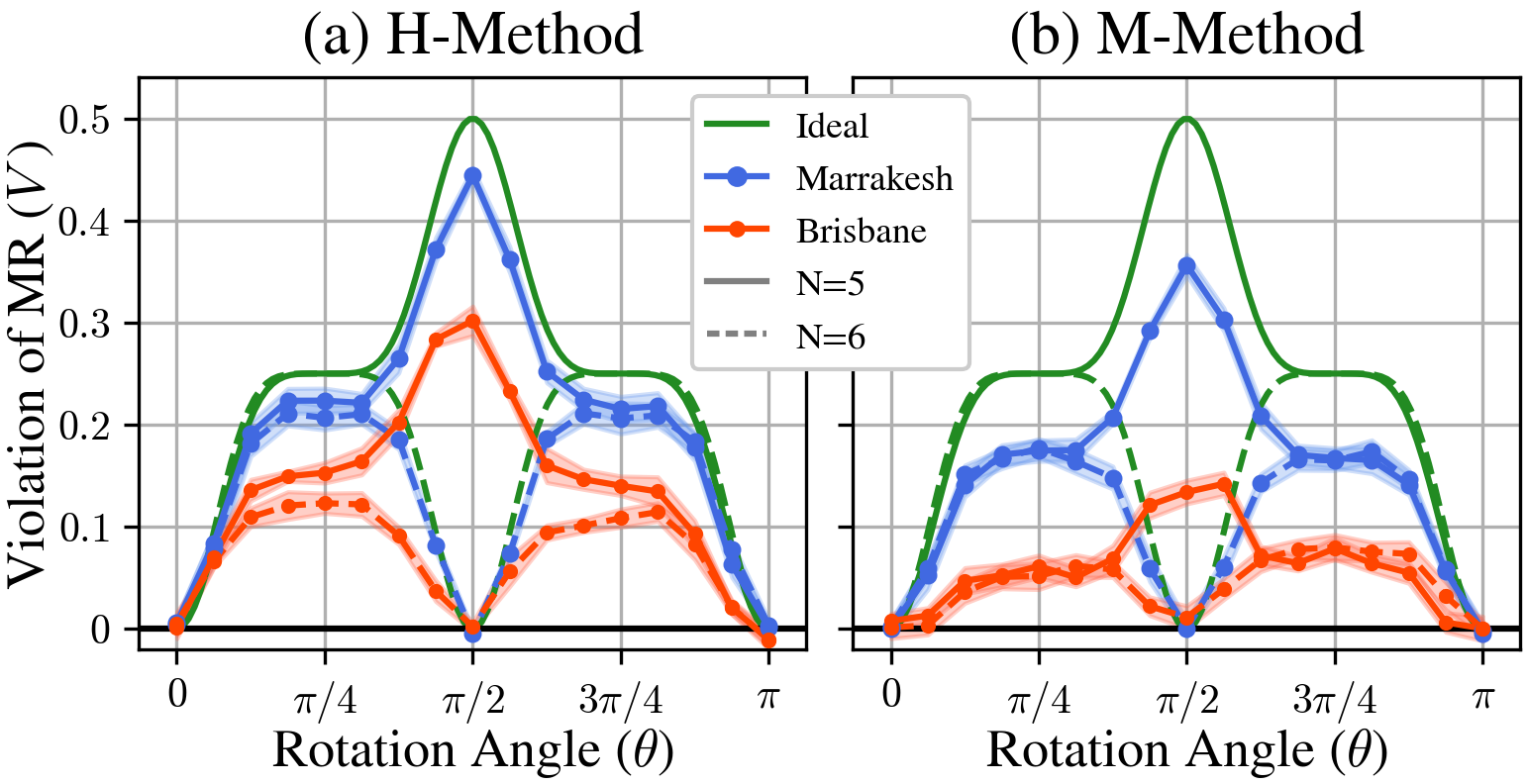}
    \caption{Violation of MR via the parity NDC protocol as a function of the rotation angle $\theta$ for the (a) H-Method and (b) M-Method, in the ideal case (green line), and the one detected in \texttt{ibm\_brisbane} (red line) and  \texttt{ibm\_marrakesh} (blue line) QCs.\label{fig:angles}}
\end{figure}

\begin{figure*}[t!]
      \includegraphics[width=\textwidth]{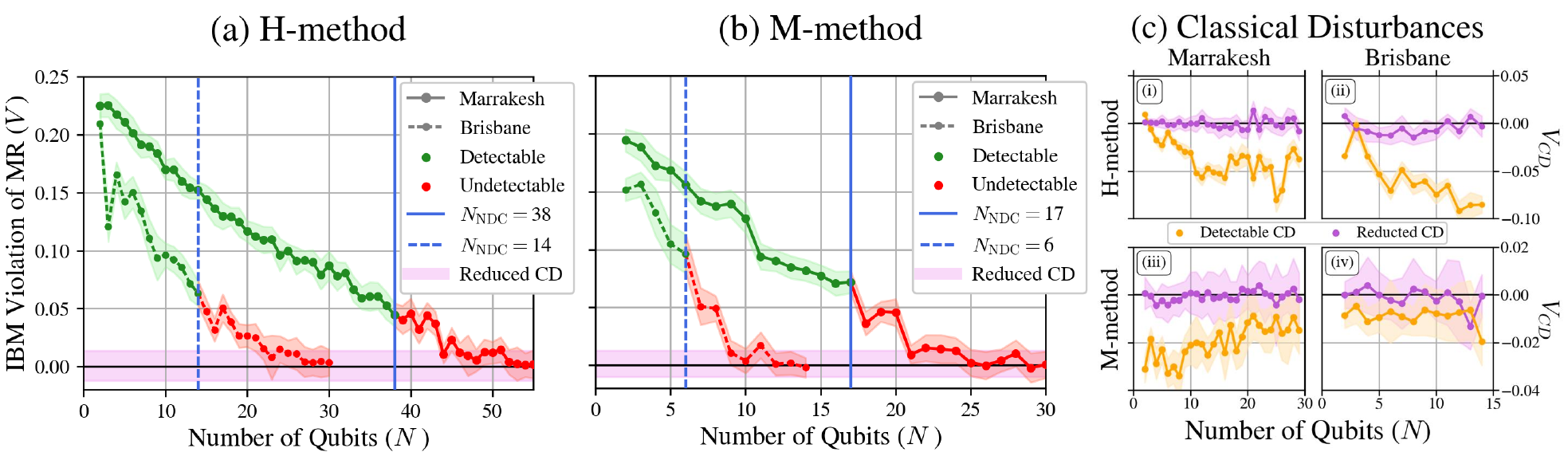}
    \caption{Violation of MR detected on IBM QCs via the parity NDC protocol as a function of the number of qubits $N$ (with $\theta=\pi/4$) implementing (a) the H-method and (b) the M-method, both used to benchmark \texttt{ibm\_brisbane} and \texttt{ibm\_marrakesh} QCs; The methods are clumsy-loophole free on both QCs as shown in (c), with $\theta=\pi$, as the detected unmitigated CDs (orange line) was reduced to zero within statistical error (purple line).}
    \label{fig:IBM_results}
\end{figure*}

In this section, we present MR violation via testing NDC for IBM QCs using the two aforementioned methods. The NDC benchmarking metric is implemented on \texttt{ibm\_brisbane} and \texttt{ibm\_marrakesh} QCs~\cite{IBM}. These computers belong to two different generations (processor types), \texttt{Eagle r3} from 2022 and \texttt{Heron r2} from 2024, respectively, showing improvements in coherence times and gate errors from the earlier to the more recent generation. Each data point reported is the average of $20$ runs, with the error bar given by its standard deviation. Each run has $n_s = 4000$ shots -- chosen so that the shot noise is smaller compared to other sources of noise measured. Error mitigation was not performed for any of the results presented below, in order to benchmark the nascent quality of the hardware.

First of all, the predicted behaviour of the NDC violation $V$ as a function of the rotation angle $\theta$ (given by Eq.~\ref{eq:violation_ideal}) is detected on both QCs with both methods, as shown in Fig. \ref{fig:angles}. Specifically, for $N=6$ (example of even number of qubits), $V$ reaches its maxima at $\theta \approx \{ \pi/4, 3 \pi/4\}$ and its minima at $\theta \approx \{ 0, \pi/2, \pi\}$, while, for $N=5$ (example of odd number of qubits), the maximum $V$ is for $\theta \approx \pi$, with the minima at $\theta \approx \{ 0, \pi\}$ and two saddle points at $\theta \approx \{ \pi/4, 3 \pi/4\}$. Compared to the ideal case, the noise present in actual implementations decreases the magnitude of the NDC violation for all measured $\theta$. The M-method -- with a higher circuit depth and run time relative to the H-method --  has a lower NDC violation. 

Crucially, for both methods, at the minima of $V$, the measured violation of the NDC is zero within the statistical error. In fact, any deviation from this value can be attributed to unwanted measurement-induced disturbances, i.e. CDs., as $V$ is predicted to be zero in the ideal quantum case (see Eq.\ref{eq:violation_ideal}). The NDC violation for $\theta = \pi$ is chosen as the control for CD ($V_{\text{CD}}:= V(\theta = \pi)$), which will be used as the reference value to claim non-classicality in the following discussion. This choice is specifically motivated by the following: (a) for both even-$N$ and odd-$N$ cases, $V=0$ for $\theta = \pi$ in the ideal quantum case; (b) it is operationally equivalent to the parity NDC protocol for $\theta = \pi/4$ (as for $\theta = 0$ the rotations are trivially removed); (c) the CDs are found to be independent of $\theta$: intuitively, the state preparation is equivalent for both sub-protocols, and CDs are induced by the difference in performing or not the intermediate measurement;
 and (d) for $\theta = \pi$, no superposition is present and hence the presented protocols are effectively ``classical computations with quantum bits"~\cite{ antonio2015classicalcomputationquantumbits}.


In Fig.~\ref{fig:IBM_results}, the main results of this work are presented certifying the quantum-to-classical transition of QCs as they become macroscopic: the detected NDC violation is inversely proportional to the number of qubits $N$ used in the protocol. In fact, although in the ideal scenario the violation of MR is found to be independent of $N$ under the parity NDC protocol (for $\theta = \pi/4$), when considering real experimental scenarios (such as the case of a QC), classical noise, decoherence, and inhomogeneities between qubits spoil the violation of NDC and the system starts behaving classically ($V \to 0$). In the literature, the tendency of a quantum system to become classical with increasing noise and macroscopicity is often referred to as the quantum-to-classical transition~\cite{large_spin1,schlosshauer_quantum_2007,MassMR2024}.
The natural phenomenon of quantum systems to become classical provides a well-motivated benchmarking protocol that measures the \textit{quantumness} of a QC as it becomes more macroscopic -- i.e., with increasing values of $N$ -- via the proposed metric, named and formally given by

\vspace{0.2cm}
\textbf{NDC Performance Metric:} \textit{is the largest number of qubits $N_{\text{NDC}}$ of a given QC for which violation of MR is detected under the parity NDC protocol without CDs, i.e., the measured NDC violation differs from the detected CDs by at least three standard deviations:}
\begin{equation}
\label{eq:discriminant}
     N_{\text{NDC}} = \max \left\{ N \; \bigg| V_{\text{NDC}}  - |V_{\text{CD}} | \geq 3 \cdot \sqrt{\sigma_{\text{NDC}}^2 + \sigma_{\text{CD}}^2} \right\} \nonumber
\end{equation}
where, for a given $N$, $V_{\text{NDC}}$ is the NDC violation ($\theta = \pi/4$) and $V_{\text{CD}}$ is the associated CD ($\theta = \pi$), both detected with uncertainties $\sigma_{\text{NDC}}$ and $\sigma_{\text{CD}}$. This implies that the behaviour of the $N$ qubits cannot be described by a classical system with $99.7\%$ confidence (under the assumption of Gaussian noise). 

The NDC Benchmarking Metric computed with \texttt{ibm\_marrakesh} are  $N_{\text{NDC}}^H =38$ and $N_{\text{NDC}}^M = 17$, and, with \texttt{ibm\_brisbane}, $N_{\text{NDC}}^H = 14$ and $N_{\text{NDC}}^M = 6$, where the superscrip $H$ ($M$) refers to the H-Method (M-method). One may note that the ratios between the two QC metrics for the two methods are similar, namely, $R^H = 38/14 \sim 2.7$ and $R^M = 17/6 \sim 2.8$, supporting the consistency of both methods. 

It is possible to conclude that both methods are clumsiness-loophole free as the CDs -- reduced using the method described in the previous section -- are detected to be zero within statistical error (Fig.~\ref{fig:IBM_results}). In fact, the detected values of CDs ($V_{\text{CD}}$) averaged over all measured $N$ are given by: $\braket{V_{\text{CD}}^H} = -0.0008 \pm 0.0105$ and $\braket{V_{\text{CD}}^M} = - 0.0004 \pm 0.0096$, for \texttt{ibm\_marrakesh}, and $\braket{V_{\text{CD}}^H} = - 0.0008 \pm 0.0126$ and $\braket{V_{\text{CD}}^M} = - 0.0011 \pm 0.0112$, for \texttt{ibm\_brisbane}. From Fig.~\ref{fig:IBM_results}, it is possible to claim loophole-freedom for (approximately) all data points, given a specific $N$. 

One may further note that, if these methods were not applied, i.e. $\hat{U}_p$ is removed entirely (mitigated via the H-method) or the mid-circuit measurement is not performed on another uncorrelated qubit (mitigated via the M-method), see Appendix~\ref{app:Classical_Disturbance}, CDs represented a significant biased noise, with averages $\braket{V_{\text{CD}}^H} = -0.037 \pm 0.020$, $\braket{V_{\text{CD}}^M} = -0.019 \pm 0.011$, and $\braket{V_{\text{CD}}^H} = -0.025 \pm 0.033$, $\braket{V_{\text{CD}}^M} = -0.009 \pm 0.010$, for \texttt{ibm\_marrakesh} and \texttt{ibm\_brisbane}, respectively. Specifically, for the former method, CDs increase with $N$ (Fig.~\ref{fig:IBM_results}.(c) - (i),(ii)) as the decomposition of $\hat{U}_p$ increases in size (and hence the difference in noise between the sub-protocols). For the latter method, CDs decrease with $N$ in the case of \texttt{ibm\_marrakesh} (Fig.~\ref{fig:IBM_results}.(c) - (iii)), for which the measurement time is 38 times longer than performing the entangling gate (but high measurement accuracy)~\cite{IBM}. This leads to higher CDs compared to \texttt{ibm\_brisbane} (Fig.~\ref{fig:IBM_results}.(c) - (iv)), for which the measurement and gate times are similar and for which the dependency on $N$ is not found~\cite{IBM}. Although the measurement on \texttt{ibm\_marrakesh} is more accurate at the single qubit level, its long execution time induces high noise in the mid-circuit parity measurement on the $N$-qubits (due, for instance, to dephasing). In the case of \texttt{ibm\_marrakesh}, CDs decreases with $N$ as the high noise due to mid-circuit parity measurement becomes comparable to the entangling gates as $N$ increases, while, conversely, for \texttt{ibm\_brisbane}, the readout noise is comparable to the entangling one already at small $N$.

\section{Conclusions \label{sec:conclusions}}

In this work, we investigate the violation of macrorealism on QCs via parity measurements. This treats the whole QC as a single macroscopic quantum entity, which is thus benchmarked. By exploiting two consecutive parity measurements, the violation of NDC is predicted to be independent of the number $N$ of qubits involved in the protocol, under ideal conditions. This is due to the fact that -- alongside unitary dynamics creating quantum coherence -- the intermediate collective parity measurement of the QC irreversibly collapses its wavefunction, thus, introducing a detectable disturbance which classical mechanics cannot describe. This metric measures the quantumness of a QC: it is a minimal test of its non-classicality, which is \textit{necessary} to achieve quantum-enhancement of any algorithm. 

The transition to classicality of a QC -- detected as the number of qubits increases -- is a foundationally motivated benchmarking protocol. Its implementation has simultaneously benchmarked (a) the quantum coherence of all qubits as the system scales, (b) the quality of parity measurements, (c) the ability to perform a mid-circuit measurement, and (d)~the ability to perform universal quantum computation, as all the required gates for the Clifford+T group are involved after compilation. Thus, our methods collectively certify the necessary requirements for performing scalable, universal, and fault-tolerant quantum computation. 


To our knowledge, this metric represents the first benchmarking protocol that involves mid-circuit parity measurements. By comparing scenarios with and without an intermediate parity measurement on an arbitrary number of qubits, the protocol investigates the errors introduced in the intermediate measurement by examining deviations from the ideal NDC violation. Specifically, the M-method is of particular interest as it performs the intermediate parity measurement as a mid-circuit measurement, thereby benchmarking an essential process for error-correcting~QCs. 

We have demonstrated this through implementations on \texttt{ibm\_brisbane} and \texttt{ibm\_marrakesh}, which also highlight the reliability of the protocols by revealing performance improvements across different generations of QCs and consistency among the two methods presented. The parity NDC benchmark can be applied to other devices and physical implementations: by comparing different QCs and platforms, its implications on technological development and its relationship with fundamental QC-noises (for instance, decoherence and inhomogeneities, which are discussed in our parallel work~\cite{LargeSpinParallel}) may be statistically realised.

To conclude, the presented results are \textit{per se} an important step forward in testing MR with macroscopic quantum systems. Using the parity NDC methods, IBM QCs have shown an order of magnitude improvement in the maximum number of qubits for which a violation of MR has been detected -- surpassing the previous limit of 5 qubits \cite{LGIQC1,LGIQC2,LGIQC3,LGIQC4,LGIQC5}. Furthermore, for the first time, this work has quantitatively analysed the quantum-to-classical transition as the system of many qubits becomes more macroscopic in an \textit{experimental} setting. 
Crucially, the parity NDC protocol -- being applicable to an arbitrary number of qubits -- can be employed in future tests of MR. Because of our construction in terms of an universal set of elementary gates (Clifford+T gates), our methodology will also be applicable to benchmark large collections of logical qubits in future fault-tolerant QCs. As more advanced machines are developed, the presented protocol may help pushing the boundary between quantum and classical physics to increasingly macroscopic systems, investigating one of the most puzzling boundaries of modern science. 

\hspace{1cm}

\section{Acknowledgments}
This project was funded and supported by the UK National Quantum Computer Centre [NQCC200921], which is a UKRI Centre and part of the UK National Quantum Technologies Programme (NQTP).
We acknowledge the use of IBM Quantum services for this work. The views expressed are those of the authors, and do not reflect the official policy or position of IBM or the IBM Quantum team. We acknowledge fruitful discussions with Alessio Serafini and Zoltán Zimborás. BZ and LB work was supported by the Engineering and Physical Sciences Research Council [Grant Numbers EP/R513143/1, EP/T517793/1, and EP/R513143/1,EP/W524335/1, respectively]. DD acknowledges the Royal Society, United Kingdom, for the support through the Newton International Fellowship (No. NIF$\backslash$R1$\backslash$212007). DD and SB acknowledge financial support from EPSRC (Engineering \& Physical Sciences Research Council, United Kingdom) Grant Numbers EP/X009467/1 and EP/R029075/1 and STFC (Science and Technology Facilities Council, United Kingdom) Grant Numbers ST/W006227/1 and ST/Z510385/1.


\begin{thebibliography}{68}%
\makeatletter
\providecommand \@ifxundefined [1]{%
 \@ifx{#1\undefined}
}%
\providecommand \@ifnum [1]{%
 \ifnum #1\expandafter \@firstoftwo
 \else \expandafter \@secondoftwo
 \fi
}%
\providecommand \@ifx [1]{%
 \ifx #1\expandafter \@firstoftwo
 \else \expandafter \@secondoftwo
 \fi
}%
\providecommand \natexlab [1]{#1}%
\providecommand \enquote  [1]{``#1''}%
\providecommand \bibnamefont  [1]{#1}%
\providecommand \bibfnamefont [1]{#1}%
\providecommand \citenamefont [1]{#1}%
\providecommand \href@noop [0]{\@secondoftwo}%
\providecommand \href [0]{\begingroup \@sanitize@url \@href}%
\providecommand \@href[1]{\@@startlink{#1}\@@href}%
\providecommand \@@href[1]{\endgroup#1\@@endlink}%
\providecommand \@sanitize@url [0]{\catcode `\\12\catcode `\$12\catcode `\&12\catcode `\#12\catcode `\^12\catcode `\_12\catcode `\%12\relax}%
\providecommand \@@startlink[1]{}%
\providecommand \@@endlink[0]{}%
\providecommand \url  [0]{\begingroup\@sanitize@url \@url }%
\providecommand \@url [1]{\endgroup\@href {#1}{\urlprefix }}%
\providecommand \urlprefix  [0]{URL }%
\providecommand \Eprint [0]{\href }%
\providecommand \doibase [0]{https://doi.org/}%
\providecommand \selectlanguage [0]{\@gobble}%
\providecommand \bibinfo  [0]{\@secondoftwo}%
\providecommand \bibfield  [0]{\@secondoftwo}%
\providecommand \translation [1]{[#1]}%
\providecommand \BibitemOpen [0]{}%
\providecommand \bibitemStop [0]{}%
\providecommand \bibitemNoStop [0]{.\EOS\space}%
\providecommand \EOS [0]{\spacefactor3000\relax}%
\providecommand \BibitemShut  [1]{\csname bibitem#1\endcsname}%
\let\auto@bib@innerbib\@empty
\bibitem [{\citenamefont {Feynman}\ \emph {et~al.}(1965)\citenamefont {Feynman} \emph {et~al.}}]{feynman_lecture_3}%
  \BibitemOpen
  \bibfield  {author} {\bibinfo {author} {\bibfnamefont {R.~P.}\ \bibnamefont {Feynman}} \emph {et~al.},\ }\href@noop {} {\emph {\bibinfo {title} {The Feynman Lectures on Physics: Quantum Mechanics}}},\ Vol.~\bibinfo {volume} {3}\ (\bibinfo  {publisher} {Addison–Wesley},\ \bibinfo {year} {1965})\BibitemShut {NoStop}%
\bibitem [{\citenamefont {Benioff}(1980)}]{Benioff_1980}%
  \BibitemOpen
  \bibfield  {author} {\bibinfo {author} {\bibfnamefont {P.}~\bibnamefont {Benioff}},\ }\bibfield  {title} {\bibinfo {title} {The computer as a physical system: A microscopic quantum mechanical hamiltonian model of computers as represented by turing machines.},\ }\href@noop {} {\bibfield  {journal} {\bibinfo  {journal} {J. Stat. Phys.}\ }\textbf {\bibinfo {volume} {22}} (\bibinfo {year} {1980})},\ \bibinfo {note} {\href{https://doi.org/10.1007/BF01011339}{10.1007/BF01011339}}\BibitemShut {NoStop}%
\bibitem [{\citenamefont {Feynman}(1982)}]{Feynman_1982}%
  \BibitemOpen
  \bibfield  {author} {\bibinfo {author} {\bibfnamefont {R.~P.}\ \bibnamefont {Feynman}},\ }\bibfield  {title} {\bibinfo {title} {Simulating physics with computer},\ }\href@noop {} {\bibfield  {journal} {\bibinfo  {journal} {Int. J. Theor. Phys}\ }\textbf {\bibinfo {volume} {21}} (\bibinfo {year} {1982})},\ \bibinfo {note} {\href{https://s2.smu.edu/~mitch/class/5395/papers/feynman-quantum-1981.pdf}{VoL21,Nos.6/7, 1982}}\BibitemShut {NoStop}%
\bibitem [{\citenamefont {Shor}(1994)}]{Shor_1994}%
  \BibitemOpen
  \bibfield  {author} {\bibinfo {author} {\bibfnamefont {P.~W.}\ \bibnamefont {Shor}},\ }\bibfield  {title} {\bibinfo {title} {Algorithms for quantum computation: discrete logarithms and factoring},\ }\href@noop {} {\bibfield  {journal} {\bibinfo  {journal} {Proceedings 35th Annual Symposium on Foundations of Computer Science}\ } (\bibinfo {year} {1994})},\ \bibinfo {note} {\href{https://dl.acm.org/doi/10.1109/sfcs.1994.365700}{10.1109/SFCS.1994.365700}}\BibitemShut {NoStop}%
\bibitem [{\citenamefont {Grover}(1997)}]{Grover_1997}%
  \BibitemOpen
  \bibfield  {author} {\bibinfo {author} {\bibfnamefont {L.~K.}\ \bibnamefont {Grover}},\ }\bibfield  {title} {\bibinfo {title} {Quantum mechanics helps in searching for a needle in a haystack},\ }\href@noop {} {\bibfield  {journal} {\bibinfo  {journal} {Phys. Rev. Lett.}\ }\textbf {\bibinfo {volume} {21}} (\bibinfo {year} {1997})},\ \bibinfo {note} {\href{https://journals.aps.org/prl/abstract/10.1103/PhysRevLett.79.325}{10.1103/PhysRevLett.79.325}}\BibitemShut {NoStop}%
\bibitem [{\citenamefont {Harrow}\ \emph {et~al.}(2009)\citenamefont {Harrow}, \citenamefont {Hassidim},\ and\ \citenamefont {Lloyd}}]{harrow_quantum_2009}%
  \BibitemOpen
  \bibfield  {author} {\bibinfo {author} {\bibfnamefont {A.~W.}\ \bibnamefont {Harrow}}, \bibinfo {author} {\bibfnamefont {A.}~\bibnamefont {Hassidim}},\ and\ \bibinfo {author} {\bibfnamefont {S.}~\bibnamefont {Lloyd}},\ }\bibfield  {title} {\bibinfo {title} {Quantum algorithm for linear systems of equations},\ }\href@noop {} {\bibfield  {journal} {\bibinfo  {journal} {Phys. Rev. Lett.}\ }\textbf {\bibinfo {volume} {103}},\ \bibinfo {pages} {150502} (\bibinfo {year} {2009})},\ \bibinfo {note} {\href{https://link.aps.org/doi/10.1103/PhysRevLett.103.150502}{10.1103/PhysRevLett.103.150502}}\BibitemShut {NoStop}%
\bibitem [{\citenamefont {Nielsen}\ and\ \citenamefont {Chuang}(2010)}]{nielsen_quantum_2010}%
  \BibitemOpen
  \bibfield  {author} {\bibinfo {author} {\bibfnamefont {M.~A.}\ \bibnamefont {Nielsen}}\ and\ \bibinfo {author} {\bibfnamefont {I.~L.}\ \bibnamefont {Chuang}},\ }\href@noop {} {\emph {\bibinfo {title} {Quantum Computation and Quantum Information: 10th Anniversary Edition}}}\ (\bibinfo  {publisher} {Cambridge University Press},\ \bibinfo {address} {Cambridge},\ \bibinfo {year} {2010})\BibitemShut {NoStop}%
\bibitem [{\citenamefont {Schmidhuber}\ \emph {et~al.}(2025)\citenamefont {Schmidhuber}, \citenamefont {O'Donnell}, \citenamefont {Kothari},\ and\ \citenamefont {Babbush}}]{Schmidhuber_Quartic}%
  \BibitemOpen
  \bibfield  {author} {\bibinfo {author} {\bibfnamefont {A.}~\bibnamefont {Schmidhuber}}, \bibinfo {author} {\bibfnamefont {R.}~\bibnamefont {O'Donnell}}, \bibinfo {author} {\bibfnamefont {R.}~\bibnamefont {Kothari}},\ and\ \bibinfo {author} {\bibfnamefont {R.}~\bibnamefont {Babbush}},\ }\bibfield  {title} {\bibinfo {title} {Quartic quantum speedups for planted inference},\ }\bibfield  {journal} {\bibinfo  {journal} {Phys. Rev. X}\ }\textbf {\bibinfo {volume} {15}},\ \href {https://doi.org/10.1103/PhysRevX.15.021077} {10.1103/PhysRevX.15.021077} (\bibinfo {year} {2025})\BibitemShut {NoStop}%
\bibitem [{\citenamefont {Knill}\ and\ \citenamefont {Laflamme}(1997)}]{knill_theory_1997}%
  \BibitemOpen
  \bibfield  {author} {\bibinfo {author} {\bibfnamefont {E.}~\bibnamefont {Knill}}\ and\ \bibinfo {author} {\bibfnamefont {R.}~\bibnamefont {Laflamme}},\ }\bibfield  {title} {\bibinfo {title} {Theory of quantum error-correcting codes},\ }\href@noop {} {\bibfield  {journal} {\bibinfo  {journal} {Phys. Rev. A}\ }\textbf {\bibinfo {volume} {55}},\ \bibinfo {pages} {900} (\bibinfo {year} {1997})},\ \bibinfo {note} {\href{https://link.aps.org/doi/10.1103/PhysRevA.55.900}{10.1103/PhysRevA.55.900}}\BibitemShut {NoStop}%
\bibitem [{\citenamefont {Gottesman}(1997)}]{gottesman_stabilizer_1997}%
  \BibitemOpen
  \bibfield  {author} {\bibinfo {author} {\bibfnamefont {D.}~\bibnamefont {Gottesman}},\ }\emph {\bibinfo {title} {Stabilizer Codes and Quantum Error Correction}},\ \href@noop {} {Ph.D. thesis},\ \bibinfo  {school} {California Institute of Technology} (\bibinfo {year} {1997}),\ \bibinfo {note} {phD thesis}\BibitemShut {NoStop}%
\bibitem [{\citenamefont {Chiaverini}\ \emph {et~al.}(2004)\citenamefont {Chiaverini}, \citenamefont {Leibfried}, \citenamefont {Schaetz}, \citenamefont {Barrett}, \citenamefont {Blakestad}, \citenamefont {Britton}, \citenamefont {Itano}, \citenamefont {Jost}, \citenamefont {Knill}, \citenamefont {Langer}, \citenamefont {Ozeri},\ and\ \citenamefont {Wineland}}]{chiaverini_realization_2004}%
  \BibitemOpen
  \bibfield  {author} {\bibinfo {author} {\bibfnamefont {J.}~\bibnamefont {Chiaverini}}, \bibinfo {author} {\bibfnamefont {D.}~\bibnamefont {Leibfried}}, \bibinfo {author} {\bibfnamefont {T.}~\bibnamefont {Schaetz}}, \bibinfo {author} {\bibfnamefont {M.~D.}\ \bibnamefont {Barrett}}, \bibinfo {author} {\bibfnamefont {R.~B.}\ \bibnamefont {Blakestad}}, \bibinfo {author} {\bibfnamefont {J.}~\bibnamefont {Britton}}, \bibinfo {author} {\bibfnamefont {W.~M.}\ \bibnamefont {Itano}}, \bibinfo {author} {\bibfnamefont {J.~D.}\ \bibnamefont {Jost}}, \bibinfo {author} {\bibfnamefont {E.}~\bibnamefont {Knill}}, \bibinfo {author} {\bibfnamefont {C.}~\bibnamefont {Langer}}, \bibinfo {author} {\bibfnamefont {R.}~\bibnamefont {Ozeri}},\ and\ \bibinfo {author} {\bibfnamefont {D.~J.}\ \bibnamefont {Wineland}},\ }\bibfield  {title} {\bibinfo {title} {Realization of quantum error correction},\ }\href@noop {} {\bibfield  {journal} {\bibinfo  {journal} {Nature}\ }\textbf {\bibinfo {volume} {432}},\ \bibinfo {pages} {602} (\bibinfo
  {year} {2004})},\ \bibinfo {note} {\href{https://doi.org/10.1038/nature03074}{10.1038/nature03074}}\BibitemShut {NoStop}%
\bibitem [{\citenamefont {Singh}\ \emph {et~al.}(2023)\citenamefont {Singh}, \citenamefont {Sundaresan}, \citenamefont {Xu}, \citenamefont {Peropadre}, \citenamefont {Jurcevic},\ and\ \citenamefont {Gambetta}}]{singh_mid-circuit_2023}%
  \BibitemOpen
  \bibfield  {author} {\bibinfo {author} {\bibfnamefont {K.}~\bibnamefont {Singh}}, \bibinfo {author} {\bibfnamefont {N.}~\bibnamefont {Sundaresan}}, \bibinfo {author} {\bibfnamefont {H.}~\bibnamefont {Xu}}, \bibinfo {author} {\bibfnamefont {B.}~\bibnamefont {Peropadre}}, \bibinfo {author} {\bibfnamefont {P.}~\bibnamefont {Jurcevic}},\ and\ \bibinfo {author} {\bibfnamefont {J.~M.}\ \bibnamefont {Gambetta}},\ }\bibfield  {title} {\bibinfo {title} {Mid-circuit correction of correlated phase errors using an array of spectator qubits},\ }\href@noop {} {\bibfield  {journal} {\bibinfo  {journal} {Science}\ }\textbf {\bibinfo {volume} {380}},\ \bibinfo {pages} {1265} (\bibinfo {year} {2023})},\ \bibinfo {note} {\href{https://doi.org/10.1126/science.ade5337}{10.1126/science.ade5337}}\BibitemShut {NoStop}%
\bibitem [{\citenamefont {{Google Quantum AI}}(2023)}]{google_quantum_ai_2023}%
  \BibitemOpen
  \bibfield  {author} {\bibinfo {author} {\bibnamefont {{Google Quantum AI}}},\ }\bibfield  {title} {\bibinfo {title} {Suppressing quantum errors by scaling a surface code logical qubit},\ }\href {https://doi.org/10.1038/s41586-022-05434-1} {\bibfield  {journal} {\bibinfo  {journal} {Nature}\ }\textbf {\bibinfo {volume} {614}},\ \bibinfo {pages} {676} (\bibinfo {year} {2023})},\ \bibinfo {note} {\href{https://doi.org/10.1038/s41586-022-05434-1}{10.1038/s41586-022-05434-1}}\BibitemShut {NoStop}%
\bibitem [{\citenamefont {Bennett}\ \emph {et~al.}(1993)\citenamefont {Bennett}, \citenamefont {Brassard} \emph {et~al.}}]{teleportation_bennett_1993}%
  \BibitemOpen
  \bibfield  {author} {\bibinfo {author} {\bibfnamefont {C.~H.}\ \bibnamefont {Bennett}}, \bibinfo {author} {\bibfnamefont {G.}~\bibnamefont {Brassard}}, \emph {et~al.},\ }\bibfield  {title} {\bibinfo {title} {Teleporting an unknown quantum state via dual classical and einstein-podolsky-rosen channels},\ }\href@noop {} {\bibfield  {journal} {\bibinfo  {journal} {Phys. Rev. Lett.}\ }\textbf {\bibinfo {volume} {70}},\ \bibinfo {pages} {1895} (\bibinfo {year} {1993})},\ \bibinfo {note} {\href{https://journals.aps.org/prl/abstract/10.1103/PhysRevLett.70.1895}{10.1103/PhysRevLett.70.1895}}\BibitemShut {NoStop}%
\bibitem [{\citenamefont {Bouwmeester}\ \emph {et~al.}(1997)\citenamefont {Bouwmeester}, \citenamefont {Pan}, \citenamefont {Mattle} \emph {et~al.}}]{bouwmeester_etal_1997}%
  \BibitemOpen
  \bibfield  {author} {\bibinfo {author} {\bibfnamefont {D.}~\bibnamefont {Bouwmeester}}, \bibinfo {author} {\bibfnamefont {J.-W.}\ \bibnamefont {Pan}}, \bibinfo {author} {\bibfnamefont {K.}~\bibnamefont {Mattle}}, \emph {et~al.},\ }\bibfield  {title} {\bibinfo {title} {Experimental quantum teleportation},\ }\href {https://doi.org/10.1038/37539} {\bibfield  {journal} {\bibinfo  {journal} {Nature}\ }\textbf {\bibinfo {volume} {390}},\ \bibinfo {pages} {575} (\bibinfo {year} {1997})},\ \bibinfo {note} {\href{https://doi.org/10.1038/37539}{10.1038/37539}}\BibitemShut {NoStop}%
\bibitem [{\citenamefont {Gottesman}\ and\ \citenamefont {Chuang}(1999)}]{gottesman_chuang_1999}%
  \BibitemOpen
  \bibfield  {author} {\bibinfo {author} {\bibfnamefont {D.}~\bibnamefont {Gottesman}}\ and\ \bibinfo {author} {\bibfnamefont {I.~L.}\ \bibnamefont {Chuang}},\ }\bibfield  {title} {\bibinfo {title} {Demonstrating the viability of universal quantum computation using teleportation and single-qubit operations},\ }\href {https://doi.org/10.1038/46503} {\bibfield  {journal} {\bibinfo  {journal} {Nature}\ }\textbf {\bibinfo {volume} {402}},\ \bibinfo {pages} {390} (\bibinfo {year} {1999})},\ \bibinfo {note} {\href{https://doi.org/10.1038/46503}{10.1038/46503}}\BibitemShut {NoStop}%
\bibitem [{\citenamefont {Kimble}(2008)}]{kimble_quantum_2008}%
  \BibitemOpen
  \bibfield  {author} {\bibinfo {author} {\bibfnamefont {H.~J.}\ \bibnamefont {Kimble}},\ }\bibfield  {title} {\bibinfo {title} {The quantum internet},\ }\href@noop {} {\bibfield  {journal} {\bibinfo  {journal} {Nature}\ }\textbf {\bibinfo {volume} {453}},\ \bibinfo {pages} {1023} (\bibinfo {year} {2008})},\ \bibinfo {note} {\href{https://doi.org/10.1038/nature07127}{10.1038/nature07127}}\BibitemShut {NoStop}%
\bibitem [{\citenamefont {B\"aumer}\ and\ \citenamefont {Woerner}(2025)}]{measurement_baumer_2025}%
  \BibitemOpen
  \bibfield  {author} {\bibinfo {author} {\bibfnamefont {E.}~\bibnamefont {B\"aumer}}\ and\ \bibinfo {author} {\bibfnamefont {S.}~\bibnamefont {Woerner}},\ }\bibfield  {title} {\bibinfo {title} {Measurement-based long-range entangling gates in constant depth},\ }\href@noop {} {\bibfield  {journal} {\bibinfo  {journal} {Phys. Rev. Res.}\ }\textbf {\bibinfo {volume} {7}},\ \bibinfo {pages} {023120} (\bibinfo {year} {2025})},\ \bibinfo {note} {\href{https://link.aps.org/doi/10.1103/PhysRevResearch.7.023120}{10.1103/PhysRevResearch.7.023120}}\BibitemShut {NoStop}%
\bibitem [{\citenamefont {Raussendorf}\ and\ \citenamefont {Briegel}(2001)}]{raussendorf_one-way_2001}%
  \BibitemOpen
  \bibfield  {author} {\bibinfo {author} {\bibfnamefont {R.}~\bibnamefont {Raussendorf}}\ and\ \bibinfo {author} {\bibfnamefont {H.~J.}\ \bibnamefont {Briegel}},\ }\bibfield  {title} {\bibinfo {title} {A one-way quantum computer},\ }\href@noop {} {\bibfield  {journal} {\bibinfo  {journal} {Phys. Rev. Lett.}\ }\textbf {\bibinfo {volume} {86}},\ \bibinfo {pages} {5188} (\bibinfo {year} {2001})},\ \bibinfo {note} {\href{https://link.aps.org/doi/10.1103/PhysRevLett.86.5188}{10.1103/PhysRevLett.86.5188}}\BibitemShut {NoStop}%
\bibitem [{\citenamefont {Raussendorf}\ \emph {et~al.}(2003)\citenamefont {Raussendorf}, \citenamefont {Browne},\ and\ \citenamefont {Briegel}}]{Raussendorf_measurement_2003}%
  \BibitemOpen
  \bibfield  {author} {\bibinfo {author} {\bibfnamefont {R.}~\bibnamefont {Raussendorf}}, \bibinfo {author} {\bibfnamefont {D.~E.}\ \bibnamefont {Browne}},\ and\ \bibinfo {author} {\bibfnamefont {H.~J.}\ \bibnamefont {Briegel}},\ }\bibfield  {title} {\bibinfo {title} {Measurement-based quantum computation on cluster states},\ }\bibfield  {journal} {\bibinfo  {journal} {Phys. Rev. A}\ }\textbf {\bibinfo {volume} {68}},\ \href {https://doi.org/10.1103/PhysRevA.68.022312} {10.1103/PhysRevA.68.022312} (\bibinfo {year} {2003})\BibitemShut {NoStop}%
\bibitem [{\citenamefont {Iqbal}\ \emph {et~al.}(2024)\citenamefont {Iqbal}, \citenamefont {Tantivasadakarn}, \citenamefont {Gatterman} \emph {et~al.}}]{Iqbal_Topological_2024}%
  \BibitemOpen
  \bibfield  {author} {\bibinfo {author} {\bibfnamefont {M.}~\bibnamefont {Iqbal}}, \bibinfo {author} {\bibfnamefont {N.}~\bibnamefont {Tantivasadakarn}}, \bibinfo {author} {\bibfnamefont {T.}~\bibnamefont {Gatterman}}, \emph {et~al.},\ }\bibfield  {title} {\bibinfo {title} {Topological order from measurements and feed-forward on a trapped ion quantum computer.},\ }\bibfield  {journal} {\bibinfo  {journal} {Commun Phys}\ }\textbf {\bibinfo {volume} {7}},\ \href {https://doi.org/10.1038/s42005-024-01698-3} {10.1038/s42005-024-01698-3} (\bibinfo {year} {2024})\BibitemShut {NoStop}%
\bibitem [{\citenamefont {Bultink}\ \emph {et~al.}(2020)\citenamefont {Bultink}, \citenamefont {O’Brien}, \citenamefont {Vollmer}, \citenamefont {Muthusubramanian}, \citenamefont {Beekman}, \citenamefont {Rol}, \citenamefont {Fu}, \citenamefont {Tarasinski}, \citenamefont {Ostroukh}, \citenamefont {Varbanov}, \citenamefont {Bruno},\ and\ \citenamefont {DiCarlo}}]{bultink_protecting_2020}%
  \BibitemOpen
  \bibfield  {author} {\bibinfo {author} {\bibfnamefont {C.~C.}\ \bibnamefont {Bultink}}, \bibinfo {author} {\bibfnamefont {T.~E.}\ \bibnamefont {O’Brien}}, \bibinfo {author} {\bibfnamefont {R.}~\bibnamefont {Vollmer}}, \bibinfo {author} {\bibfnamefont {N.}~\bibnamefont {Muthusubramanian}}, \bibinfo {author} {\bibfnamefont {M.~W.}\ \bibnamefont {Beekman}}, \bibinfo {author} {\bibfnamefont {M.~A.}\ \bibnamefont {Rol}}, \bibinfo {author} {\bibfnamefont {X.}~\bibnamefont {Fu}}, \bibinfo {author} {\bibfnamefont {B.}~\bibnamefont {Tarasinski}}, \bibinfo {author} {\bibfnamefont {V.}~\bibnamefont {Ostroukh}}, \bibinfo {author} {\bibfnamefont {B.}~\bibnamefont {Varbanov}}, \bibinfo {author} {\bibfnamefont {A.}~\bibnamefont {Bruno}},\ and\ \bibinfo {author} {\bibfnamefont {L.}~\bibnamefont {DiCarlo}},\ }\bibfield  {title} {\bibinfo {title} {Protecting quantum entanglement from leakage and qubit errors via repetitive parity measurements},\ }\href@noop {} {\bibfield  {journal} {\bibinfo  {journal} {Science Advances}\
  }\textbf {\bibinfo {volume} {6}} (\bibinfo {year} {2020})},\ \bibinfo {note} {\href{https://www.science.org/doi/10.1126/sciadv.aay3050}{10.1126/sciadv.aay3050}}\BibitemShut {NoStop}%
\bibitem [{\citenamefont {Smith}\ \emph {et~al.}(2025)\citenamefont {Smith}, \citenamefont {Klaver}, \citenamefont {Nautrup}, \citenamefont {Lechner},\ and\ \citenamefont {Briegel}}]{smith_minimally_2025}%
  \BibitemOpen
  \bibfield  {author} {\bibinfo {author} {\bibfnamefont {I.~D.}\ \bibnamefont {Smith}}, \bibinfo {author} {\bibfnamefont {B.}~\bibnamefont {Klaver}}, \bibinfo {author} {\bibfnamefont {H.~P.}\ \bibnamefont {Nautrup}}, \bibinfo {author} {\bibfnamefont {W.}~\bibnamefont {Lechner}},\ and\ \bibinfo {author} {\bibfnamefont {H.~J.}\ \bibnamefont {Briegel}},\ }\href {http://arxiv.org/abs/2504.03556} {\bibinfo {title} {Minimally {Universal} {Parity} {Quantum} {Computing}}} (\bibinfo {year} {2025}),\ \Eprint {https://arxiv.org/abs/2504.03556} {arXiv:2504.03556 [quant-ph]} \BibitemShut {NoStop}%
\bibitem [{\citenamefont {Fellner}\ \emph {et~al.}(2022)\citenamefont {Fellner}, \citenamefont {Messinger}, \citenamefont {Ender},\ and\ \citenamefont {Lechner}}]{fellner_universal_2022}%
  \BibitemOpen
  \bibfield  {author} {\bibinfo {author} {\bibfnamefont {M.}~\bibnamefont {Fellner}}, \bibinfo {author} {\bibfnamefont {A.}~\bibnamefont {Messinger}}, \bibinfo {author} {\bibfnamefont {K.}~\bibnamefont {Ender}},\ and\ \bibinfo {author} {\bibfnamefont {W.}~\bibnamefont {Lechner}},\ }\bibfield  {title} {\bibinfo {title} {Universal {Parity} {Quantum} {Computing}},\ }\href@noop {} {\bibfield  {journal} {\bibinfo  {journal} {Physical Review Letters}\ }\textbf {\bibinfo {volume} {129}} (\bibinfo {year} {2022})},\ \bibinfo {note} {\href{https://link.aps.org/doi/10.1103/PhysRevLett.129.180503}{10.1103/PhysRevLett.129.180503}}\BibitemShut {NoStop}%
\bibitem [{\citenamefont {Saira}\ \emph {et~al.}(2014)\citenamefont {Saira}, \citenamefont {Groen}, \citenamefont {Cramer}, \citenamefont {Meretska}, \citenamefont {De~Lange},\ and\ \citenamefont {DiCarlo}}]{saira_entanglement_2014}%
  \BibitemOpen
  \bibfield  {author} {\bibinfo {author} {\bibfnamefont {O.-P.}\ \bibnamefont {Saira}}, \bibinfo {author} {\bibfnamefont {J.}~\bibnamefont {Groen}}, \bibinfo {author} {\bibfnamefont {J.}~\bibnamefont {Cramer}}, \bibinfo {author} {\bibfnamefont {M.}~\bibnamefont {Meretska}}, \bibinfo {author} {\bibfnamefont {G.}~\bibnamefont {De~Lange}},\ and\ \bibinfo {author} {\bibfnamefont {L.}~\bibnamefont {DiCarlo}},\ }\bibfield  {title} {\bibinfo {title} {Entanglement {Genesis} by {Ancilla}-{Based} {Parity} {Measurement} in {2D} {Circuit} {QED}},\ }\href@noop {} {\bibfield  {journal} {\bibinfo  {journal} {Physical Review Letters}\ }\textbf {\bibinfo {volume} {112}} (\bibinfo {year} {2014})},\ \bibinfo {note} {\href{https://link.aps.org/doi/10.1103/PhysRevLett.112.070502}{10.1103/PhysRevLett.112.070502}}\BibitemShut {NoStop}%
\bibitem [{\citenamefont {Lall}\ \emph {et~al.}(2025)\citenamefont {Lall}, \citenamefont {Agarwal} \emph {et~al.}}]{lall_review_2025}%
  \BibitemOpen
  \bibfield  {author} {\bibinfo {author} {\bibfnamefont {D.}~\bibnamefont {Lall}}, \bibinfo {author} {\bibfnamefont {A.}~\bibnamefont {Agarwal}}, \emph {et~al.},\ }\href {http://arxiv.org/abs/2502.06717} {\bibinfo {title} {A {Review} and {Collection} of {Metrics} and {Benchmarks} for {Quantum} {Computers}: definitions, methodologies and software}} (\bibinfo {year} {2025}),\ \Eprint {https://arxiv.org/abs/2502.06717} {arXiv:2502.06717 [quant-ph]} \BibitemShut {NoStop}%
\bibitem [{\citenamefont {Wack}\ \emph {et~al.}(2021)\citenamefont {Wack}, \citenamefont {Paik}, \citenamefont {Javadi-Abhari}, \citenamefont {Jurcevic}, \citenamefont {Faro}, \citenamefont {Gambetta},\ and\ \citenamefont {Johnson}}]{wack_quality_2021}%
  \BibitemOpen
  \bibfield  {author} {\bibinfo {author} {\bibfnamefont {A.}~\bibnamefont {Wack}}, \bibinfo {author} {\bibfnamefont {H.}~\bibnamefont {Paik}}, \bibinfo {author} {\bibfnamefont {A.}~\bibnamefont {Javadi-Abhari}}, \bibinfo {author} {\bibfnamefont {P.}~\bibnamefont {Jurcevic}}, \bibinfo {author} {\bibfnamefont {I.}~\bibnamefont {Faro}}, \bibinfo {author} {\bibfnamefont {J.~M.}\ \bibnamefont {Gambetta}},\ and\ \bibinfo {author} {\bibfnamefont {B.}~\bibnamefont {Johnson}},\ }\href@noop {} {\bibinfo {title} {Quality, speed, and scale: three key attributes to measure the performance of near-term quantum computers}} (\bibinfo {year} {2021}),\ \bibinfo {note} {\href{https://arxiv.org/abs/2110.14108}{arXiv:2110.14108}}\BibitemShut {NoStop}%
\bibitem [{\citenamefont {Cross}\ \emph {et~al.}(2019{\natexlab{a}})\citenamefont {Cross}, \citenamefont {Bishop}, \citenamefont {Sheldon}, \citenamefont {Nation},\ and\ \citenamefont {Gambetta}}]{benchmarking_volume}%
  \BibitemOpen
  \bibfield  {author} {\bibinfo {author} {\bibfnamefont {A.~W.}\ \bibnamefont {Cross}}, \bibinfo {author} {\bibfnamefont {L.~S.}\ \bibnamefont {Bishop}}, \bibinfo {author} {\bibfnamefont {S.}~\bibnamefont {Sheldon}}, \bibinfo {author} {\bibfnamefont {P.~D.}\ \bibnamefont {Nation}},\ and\ \bibinfo {author} {\bibfnamefont {J.~M.}\ \bibnamefont {Gambetta}},\ }\bibfield  {title} {\bibinfo {title} {Validating quantum computers using randomized model circuits},\ }\href {https://doi.org/10.1103/PhysRevA.100.032328} {\bibfield  {journal} {\bibinfo  {journal} {Phys. Rev. A}\ }\textbf {\bibinfo {volume} {100}},\ \bibinfo {pages} {032328} (\bibinfo {year} {2019}{\natexlab{a}})}\BibitemShut {NoStop}%
\bibitem [{\citenamefont {Baldwin}\ \emph {et~al.}(2022)\citenamefont {Baldwin}, \citenamefont {Mayer} \emph {et~al.}}]{benchmarking_volume_2}%
  \BibitemOpen
  \bibfield  {author} {\bibinfo {author} {\bibfnamefont {C.~H.}\ \bibnamefont {Baldwin}}, \bibinfo {author} {\bibfnamefont {K.}~\bibnamefont {Mayer}}, \emph {et~al.},\ }\bibfield  {title} {\bibinfo {title} {Re-examining the quantum volume test: Ideal distributions, compiler optimizations, confidence intervals, and scalable resource estimations},\ }\href {https://doi.org/10.22331/q-2022-05-09-707} {\bibfield  {journal} {\bibinfo  {journal} {Quantum}\ }\textbf {\bibinfo {volume} {6}},\ \bibinfo {pages} {707} (\bibinfo {year} {2022})}\BibitemShut {NoStop}%
\bibitem [{\citenamefont {Magesan}\ \emph {et~al.}(2011)\citenamefont {Magesan}, \citenamefont {Gambetta},\ and\ \citenamefont {Emerson}}]{magesan_scalable_2011}%
  \BibitemOpen
  \bibfield  {author} {\bibinfo {author} {\bibfnamefont {E.}~\bibnamefont {Magesan}}, \bibinfo {author} {\bibfnamefont {J.~M.}\ \bibnamefont {Gambetta}},\ and\ \bibinfo {author} {\bibfnamefont {J.}~\bibnamefont {Emerson}},\ }\bibfield  {title} {\bibinfo {title} {Scalable and robust randomized benchmarking of quantum processes},\ }\href {https://link.aps.org/doi/10.1103/PhysRevLett.106.180504} {\bibfield  {journal} {\bibinfo  {journal} {Phys. Rev. Lett.}\ }\textbf {\bibinfo {volume} {106}},\ \bibinfo {pages} {180504} (\bibinfo {year} {2011})},\ \bibinfo {note} {\href{https://link.aps.org/doi/10.1103/PhysRevLett.106.180504}{10.1103/PhysRevLett.106.180504}}\BibitemShut {NoStop}%
\bibitem [{\citenamefont {Kimmel}\ \emph {et~al.}(2014)\citenamefont {Kimmel}, \citenamefont {da~Silva}, \citenamefont {Ryan}, \citenamefont {Johnson},\ and\ \citenamefont {Ohki}}]{kimmel_robust_2014}%
  \BibitemOpen
  \bibfield  {author} {\bibinfo {author} {\bibfnamefont {S.}~\bibnamefont {Kimmel}}, \bibinfo {author} {\bibfnamefont {M.~P.}\ \bibnamefont {da~Silva}}, \bibinfo {author} {\bibfnamefont {C.~A.}\ \bibnamefont {Ryan}}, \bibinfo {author} {\bibfnamefont {B.~R.}\ \bibnamefont {Johnson}},\ and\ \bibinfo {author} {\bibfnamefont {T.}~\bibnamefont {Ohki}},\ }\bibfield  {title} {\bibinfo {title} {Robust extraction of tomographic information via randomized benchmarking},\ }\href {https://link.aps.org/doi/10.1103/PhysRevX.4.011050} {\bibfield  {journal} {\bibinfo  {journal} {Phys. Rev. X}\ }\textbf {\bibinfo {volume} {4}},\ \bibinfo {pages} {011050} (\bibinfo {year} {2014})},\ \bibinfo {note} {\href{https://link.aps.org/doi/10.1103/PhysRevX.4.011050}{10.1103/PhysRevX.4.011050}}\BibitemShut {NoStop}%
\bibitem [{\citenamefont {Boixo}\ \emph {et~al.}(2018)\citenamefont {Boixo}, \citenamefont {Isakov}, \citenamefont {Smelyanskiy},\ and\ \citenamefont {\textit{et al.}}}]{boixo_characterizing_2018}%
  \BibitemOpen
  \bibfield  {author} {\bibinfo {author} {\bibfnamefont {S.}~\bibnamefont {Boixo}}, \bibinfo {author} {\bibfnamefont {S.~V.}\ \bibnamefont {Isakov}}, \bibinfo {author} {\bibfnamefont {V.~N.}\ \bibnamefont {Smelyanskiy}},\ and\ \bibinfo {author} {\bibnamefont {\textit{et al.}}},\ }\bibfield  {title} {\bibinfo {title} {Characterizing quantum supremacy in near-term devices},\ }\href@noop {} {\bibfield  {journal} {\bibinfo  {journal} {Nature Physics}\ }\textbf {\bibinfo {volume} {14}},\ \bibinfo {pages} {595} (\bibinfo {year} {2018})},\ \bibinfo {note} {\href{https://doi.org/10.1038/s41567-018-0124-x}{10.1038/s41567-018-0124-x}}\BibitemShut {NoStop}%
\bibitem [{\citenamefont {Cross}\ \emph {et~al.}(2019{\natexlab{b}})\citenamefont {Cross}, \citenamefont {Bishop}, \citenamefont {Sheldon}, \citenamefont {Nation},\ and\ \citenamefont {Gambetta}}]{cross_validating_2019}%
  \BibitemOpen
  \bibfield  {author} {\bibinfo {author} {\bibfnamefont {A.~W.}\ \bibnamefont {Cross}}, \bibinfo {author} {\bibfnamefont {L.~S.}\ \bibnamefont {Bishop}}, \bibinfo {author} {\bibfnamefont {S.}~\bibnamefont {Sheldon}}, \bibinfo {author} {\bibfnamefont {P.~D.}\ \bibnamefont {Nation}},\ and\ \bibinfo {author} {\bibfnamefont {J.~M.}\ \bibnamefont {Gambetta}},\ }\bibfield  {title} {\bibinfo {title} {Validating quantum computers using randomized model circuits},\ }\href@noop {} {\bibfield  {journal} {\bibinfo  {journal} {Phys. Rev. A}\ }\textbf {\bibinfo {volume} {100}},\ \bibinfo {pages} {032328} (\bibinfo {year} {2019}{\natexlab{b}})},\ \bibinfo {note} {\href{https://link.aps.org/doi/10.1103/PhysRevA.100.032328}{10.1103/PhysRevA.100.032328}}\BibitemShut {NoStop}%
\bibitem [{\citenamefont {Tomesh}\ \emph {et~al.}(2022)\citenamefont {Tomesh}, \citenamefont {Gokhale}, \citenamefont {Omole},\ and\ \citenamefont {\textit{et al.}}}]{tomesh_supermarq_2022}%
  \BibitemOpen
  \bibfield  {author} {\bibinfo {author} {\bibfnamefont {T.}~\bibnamefont {Tomesh}}, \bibinfo {author} {\bibfnamefont {P.}~\bibnamefont {Gokhale}}, \bibinfo {author} {\bibfnamefont {V.}~\bibnamefont {Omole}},\ and\ \bibinfo {author} {\bibnamefont {\textit{et al.}}},\ }\href@noop {} {\bibinfo {title} {Supermarq: A scalable quantum benchmark suite}} (\bibinfo {year} {2022}),\ \bibinfo {note} {\href{https://arxiv.org/abs/2202.11045}{arXiv:2202.11045}}\BibitemShut {NoStop}%
\bibitem [{\citenamefont {Martiel}\ \emph {et~al.}(2021)\citenamefont {Martiel}, \citenamefont {Ayral},\ and\ \citenamefont {Allouche}}]{benchmarking_variation}%
  \BibitemOpen
  \bibfield  {author} {\bibinfo {author} {\bibfnamefont {S.}~\bibnamefont {Martiel}}, \bibinfo {author} {\bibfnamefont {T.}~\bibnamefont {Ayral}},\ and\ \bibinfo {author} {\bibfnamefont {C.}~\bibnamefont {Allouche}},\ }\bibfield  {title} {\bibinfo {title} {Benchmarking quantum coprocessors in an application-centric, hardware-agnostic, and scalable way},\ }\href {https://doi.org/10.1109/TQE.2021.3090207} {\bibfield  {journal} {\bibinfo  {journal} {IEEE Transactions on Quantum Engineering}\ }\textbf {\bibinfo {volume} {2}},\ \bibinfo {pages} {1} (\bibinfo {year} {2021})}\BibitemShut {NoStop}%
\bibitem [{\citenamefont {Chen}\ \emph {et~al.}(2024)\citenamefont {Chen}, \citenamefont {Nielsen}, \citenamefont {Ebert}, \citenamefont {Inlek}, \citenamefont {Wright}, \citenamefont {Chaplin}, \citenamefont {Maksymov}, \citenamefont {P{\'{a}}ez}, \citenamefont {Poudel}, \citenamefont {Maunz},\ and\ \citenamefont {Gamble}}]{benchmarking_algo}%
  \BibitemOpen
  \bibfield  {author} {\bibinfo {author} {\bibfnamefont {J.-S.}\ \bibnamefont {Chen}}, \bibinfo {author} {\bibfnamefont {E.}~\bibnamefont {Nielsen}}, \bibinfo {author} {\bibfnamefont {M.}~\bibnamefont {Ebert}}, \bibinfo {author} {\bibfnamefont {V.}~\bibnamefont {Inlek}}, \bibinfo {author} {\bibfnamefont {K.}~\bibnamefont {Wright}}, \bibinfo {author} {\bibfnamefont {V.}~\bibnamefont {Chaplin}}, \bibinfo {author} {\bibfnamefont {A.}~\bibnamefont {Maksymov}}, \bibinfo {author} {\bibfnamefont {E.}~\bibnamefont {P{\'{a}}ez}}, \bibinfo {author} {\bibfnamefont {A.}~\bibnamefont {Poudel}}, \bibinfo {author} {\bibfnamefont {P.}~\bibnamefont {Maunz}},\ and\ \bibinfo {author} {\bibfnamefont {J.}~\bibnamefont {Gamble}},\ }\bibfield  {title} {\bibinfo {title} {Benchmarking a trapped-ion quantum computer with 30 qubits},\ }\href {https://doi.org/10.22331/q-2024-11-07-1516} {\bibfield  {journal} {\bibinfo  {journal} {{Quantum}}\ }\textbf {\bibinfo {volume} {8}},\ \bibinfo {pages} {1516} (\bibinfo {year} {2024})}\BibitemShut
  {NoStop}%
\bibitem [{\citenamefont {Ferracin}\ \emph {et~al.}(2018)\citenamefont {Ferracin}, \citenamefont {Kapourniotis},\ and\ \citenamefont {Datta}}]{Ferracin2018reducing}%
  \BibitemOpen
  \bibfield  {author} {\bibinfo {author} {\bibfnamefont {S.}~\bibnamefont {Ferracin}}, \bibinfo {author} {\bibfnamefont {T.}~\bibnamefont {Kapourniotis}},\ and\ \bibinfo {author} {\bibfnamefont {A.}~\bibnamefont {Datta}},\ }\bibfield  {title} {\bibinfo {title} {Reducing resources for verification of quantum computations},\ }\href {https://doi.org/10.1103/PhysRevA.98.022323} {\bibfield  {journal} {\bibinfo  {journal} {Phys. Rev. A}\ }\textbf {\bibinfo {volume} {98}},\ \bibinfo {pages} {022323} (\bibinfo {year} {2018})}\BibitemShut {NoStop}%
\bibitem [{\citenamefont {Ferracin}\ \emph {et~al.}(2019)\citenamefont {Ferracin}, \citenamefont {Kapourniotis},\ and\ \citenamefont {Datta}}]{Ferracin2019accrediting}%
  \BibitemOpen
  \bibfield  {author} {\bibinfo {author} {\bibfnamefont {S.}~\bibnamefont {Ferracin}}, \bibinfo {author} {\bibfnamefont {T.}~\bibnamefont {Kapourniotis}},\ and\ \bibinfo {author} {\bibfnamefont {A.}~\bibnamefont {Datta}},\ }\bibfield  {title} {\bibinfo {title} {Accrediting outputs of noisy intermediate-scale quantum computing devices},\ }\href {https://doi.org/10.1088/1367-2630/ab4fd6} {\bibfield  {journal} {\bibinfo  {journal} {New Journal of Physics}\ }\textbf {\bibinfo {volume} {21}},\ \bibinfo {pages} {113038} (\bibinfo {year} {2019})}\BibitemShut {NoStop}%
\bibitem [{\citenamefont {Govia}\ \emph {et~al.}(2023)\citenamefont {Govia}, \citenamefont {Jurcevic} \emph {et~al.}}]{govia_randomized_2023}%
  \BibitemOpen
  \bibfield  {author} {\bibinfo {author} {\bibfnamefont {L.~C.~G.}\ \bibnamefont {Govia}}, \bibinfo {author} {\bibfnamefont {P.}~\bibnamefont {Jurcevic}}, \emph {et~al.},\ }\bibfield  {title} {\bibinfo {title} {A randomized benchmarking suite for mid-circuit measurements},\ }\href {https://iopscience.iop.org/article/10.1088/1367-2630/ad0e19} {\bibfield  {journal} {\bibinfo  {journal} {New J. Phys.}\ }\textbf {\bibinfo {volume} {25}} (\bibinfo {year} {2023})},\ \bibinfo {note} {\href{https://iopscience.iop.org/article/10.1088/1367-2630/ad0e19}{10.1088/1367-2630/ad0e19}}\BibitemShut {NoStop}%
\bibitem [{\citenamefont {Hothem}\ \emph {et~al.}(2024)\citenamefont {Hothem}, \citenamefont {Hines}, \citenamefont {Baldwin}, \citenamefont {Gresh}, \citenamefont {Blume-Kohout},\ and\ \citenamefont {Proctor}}]{hothem_measuring_2024}%
  \BibitemOpen
  \bibfield  {author} {\bibinfo {author} {\bibfnamefont {D.}~\bibnamefont {Hothem}}, \bibinfo {author} {\bibfnamefont {J.}~\bibnamefont {Hines}}, \bibinfo {author} {\bibfnamefont {C.}~\bibnamefont {Baldwin}}, \bibinfo {author} {\bibfnamefont {D.}~\bibnamefont {Gresh}}, \bibinfo {author} {\bibfnamefont {R.}~\bibnamefont {Blume-Kohout}},\ and\ \bibinfo {author} {\bibfnamefont {T.}~\bibnamefont {Proctor}},\ }\href {https://arxiv.org/abs/2410.16706} {\bibinfo {title} {Measuring error rates of mid-circuit measurements}} (\bibinfo {year} {2024}),\ \bibinfo {note} {\href{https://arxiv.org/abs/2410.16706}{arXiv:2410.16706}},\ \Eprint {https://arxiv.org/abs/2410.16706} {arXiv:2410.16706 [quant-ph]} \BibitemShut {NoStop}%
\bibitem [{\citenamefont {Bell}(1964)}]{Bell}%
  \BibitemOpen
  \bibfield  {author} {\bibinfo {author} {\bibfnamefont {J.~S.}\ \bibnamefont {Bell}},\ }\bibfield  {title} {\bibinfo {title} {On the einstein podolsky rosen paradox},\ }\href {https://doi.org/10.1103/PhysicsPhysiqueFizika.1.195} {\bibfield  {journal} {\bibinfo  {journal} {Physics Physique Fizika}\ }\textbf {\bibinfo {volume} {1}},\ \bibinfo {pages} {195} (\bibinfo {year} {1964})}\BibitemShut {NoStop}%
\bibitem [{\citenamefont {Leggett}\ and\ \citenamefont {Garg}(1985)}]{lgi1}%
  \BibitemOpen
  \bibfield  {author} {\bibinfo {author} {\bibfnamefont {A.~J.}\ \bibnamefont {Leggett}}\ and\ \bibinfo {author} {\bibfnamefont {A.}~\bibnamefont {Garg}},\ }\bibfield  {title} {\bibinfo {title} {Quantum mechanics versus macroscopic realism: Is the flux there when nobody looks?},\ }\href@noop {} {\bibfield  {journal} {\bibinfo  {journal} {Phys. Rev. Lett.}\ }\textbf {\bibinfo {volume} {54}} (\bibinfo {year} {1985})},\ \bibinfo {note} {\href{https://doi.org/10.1103/PhysRevLett.54.857}{10.1103/PhysRevLett.54.857}}\BibitemShut {NoStop}%
\bibitem [{\citenamefont {Leggett}(2002)}]{leggett02}%
  \BibitemOpen
  \bibfield  {author} {\bibinfo {author} {\bibfnamefont {A.~J.}\ \bibnamefont {Leggett}},\ }\bibfield  {title} {\bibinfo {title} {Testing the limits of quantum mechanics: motivation, state of play, prospects},\ }\href@noop {} {\bibfield  {journal} {\bibinfo  {journal} {Phys. Condens. Matter}\ }\textbf {\bibinfo {volume} {14}} (\bibinfo {year} {2002})},\ \bibinfo {note} {\href{https://doi.org/10.1088/0953-8984/14/15/201}{10.1088/0953-8984/14/15/201}}\BibitemShut {NoStop}%
\bibitem [{\citenamefont {Leggett}(2008)}]{lgi2}%
  \BibitemOpen
  \bibfield  {author} {\bibinfo {author} {\bibfnamefont {A.~J.}\ \bibnamefont {Leggett}},\ }\bibfield  {title} {\bibinfo {title} {Realism and the physical world},\ }\href@noop {} {\bibfield  {journal} {\bibinfo  {journal} {Rep. Prog. Phys.}\ }\textbf {\bibinfo {volume} {71}} (\bibinfo {year} {2008})},\ \bibinfo {note} {\href{https://doi.org/10.1088/0034-4885/71/2/022001}{10.1126/sciadv.abg2879}}\BibitemShut {NoStop}%
\bibitem [{\citenamefont {Emary}\ \emph {et~al.}(2014)\citenamefont {Emary}, \citenamefont {Lambert},\ and\ \citenamefont {Nori}}]{qlgi1}%
  \BibitemOpen
  \bibfield  {author} {\bibinfo {author} {\bibfnamefont {C.}~\bibnamefont {Emary}}, \bibinfo {author} {\bibfnamefont {N.}~\bibnamefont {Lambert}},\ and\ \bibinfo {author} {\bibfnamefont {F.}~\bibnamefont {Nori}},\ }\bibfield  {title} {\bibinfo {title} {Leggett–garg inequalities},\ }\href@noop {} {\bibfield  {journal} {\bibinfo  {journal} {Rep. Prog. Phys.}\ }\textbf {\bibinfo {volume} {77}} (\bibinfo {year} {2014})},\ \bibinfo {note} {\href{https://doi.org/10.1088/0034-4885/77/1/016001}{10.1088/0034-4885/77/1/016001}}\BibitemShut {NoStop}%
\bibitem [{\citenamefont {Knee}\ \emph {et~al.}(2016{\natexlab{a}})\citenamefont {Knee}, \citenamefont {Kakuyanagi} \emph {et~al.}}]{nsit1}%
  \BibitemOpen
  \bibfield  {author} {\bibinfo {author} {\bibfnamefont {G.~C.}\ \bibnamefont {Knee}}, \bibinfo {author} {\bibfnamefont {K.}~\bibnamefont {Kakuyanagi}}, \emph {et~al.},\ }\bibfield  {title} {\bibinfo {title} {A strict experimental test of macroscopic realism in a superconducting flux qubit},\ }\href@noop {} {\bibfield  {journal} {\bibinfo  {journal} {Nature Communications}\ }\textbf {\bibinfo {volume} {7}} (\bibinfo {year} {2016}{\natexlab{a}})},\ \bibinfo {note} {\href{https://doi.org/10.1038/ncomms13253}{10.1038/ncomms13253}}\BibitemShut {NoStop}%
\bibitem [{\citenamefont {Kofler}\ and\ \citenamefont {Brukner}(2013)}]{nsit2}%
  \BibitemOpen
  \bibfield  {author} {\bibinfo {author} {\bibfnamefont {J.}~\bibnamefont {Kofler}}\ and\ \bibinfo {author} {\bibfnamefont {C.}~\bibnamefont {Brukner}},\ }\bibfield  {title} {\bibinfo {title} {Condition for macroscopic realism beyond the leggett-garg inequalities},\ }\href@noop {} {\bibfield  {journal} {\bibinfo  {journal} {Phys. Rev. A}\ }\textbf {\bibinfo {volume} {87}} (\bibinfo {year} {2013})},\ \bibinfo {note} {\href{https://doi.org/10.1103/PhysRevA.87.052115}{10.1103/PhysRevA.87.052115}}\BibitemShut {NoStop}%
\bibitem [{\citenamefont {Schild}\ and\ \citenamefont {Emary}(2015)}]{nsit3}%
  \BibitemOpen
  \bibfield  {author} {\bibinfo {author} {\bibfnamefont {G.}~\bibnamefont {Schild}}\ and\ \bibinfo {author} {\bibfnamefont {C.}~\bibnamefont {Emary}},\ }\bibfield  {title} {\bibinfo {title} {Maximum violations of the quantum-witness equality},\ }\href@noop {} {\bibfield  {journal} {\bibinfo  {journal} {Phys. Rev. A}\ }\textbf {\bibinfo {volume} {92}} (\bibinfo {year} {2015})},\ \bibinfo {note} {\href{https://doi.org/10.1103/PhysRevA.92.032101}{10.1103/PhysRevA.92.032101}}\BibitemShut {NoStop}%
\bibitem [{\citenamefont {Hanif}\ \emph {et~al.}(2024)\citenamefont {Hanif}, \citenamefont {Das}, \citenamefont {Halliwell}, \citenamefont {Home}, \citenamefont {Mazumdar}, \citenamefont {Ulbricht},\ and\ \citenamefont {Bose}}]{gravity_measurement}%
  \BibitemOpen
  \bibfield  {author} {\bibinfo {author} {\bibfnamefont {F.}~\bibnamefont {Hanif}}, \bibinfo {author} {\bibfnamefont {D.}~\bibnamefont {Das}}, \bibinfo {author} {\bibfnamefont {J.}~\bibnamefont {Halliwell}}, \bibinfo {author} {\bibfnamefont {D.}~\bibnamefont {Home}}, \bibinfo {author} {\bibfnamefont {A.}~\bibnamefont {Mazumdar}}, \bibinfo {author} {\bibfnamefont {H.}~\bibnamefont {Ulbricht}},\ and\ \bibinfo {author} {\bibfnamefont {S.}~\bibnamefont {Bose}},\ }\bibfield  {title} {\bibinfo {title} {Testing whether gravity acts as a quantum entity when measured},\ }\href {https://doi.org/10.1103/PhysRevLett.133.180201} {\bibfield  {journal} {\bibinfo  {journal} {Phys. Rev. Lett.}\ }\textbf {\bibinfo {volume} {133}},\ \bibinfo {pages} {180201} (\bibinfo {year} {2024})}\BibitemShut {NoStop}%
\bibitem [{\citenamefont {Athalye}\ \emph {et~al.}(2011)\citenamefont {Athalye}, \citenamefont {Roy},\ and\ \citenamefont {Mahesh}}]{Athalye_2011}%
  \BibitemOpen
  \bibfield  {author} {\bibinfo {author} {\bibfnamefont {V.}~\bibnamefont {Athalye}}, \bibinfo {author} {\bibfnamefont {S.~S.}\ \bibnamefont {Roy}},\ and\ \bibinfo {author} {\bibfnamefont {T.~S.}\ \bibnamefont {Mahesh}},\ }\bibfield  {title} {\bibinfo {title} {Investigation of the leggett-garg inequality for precessing nuclear spins},\ }\href {https://doi.org/10.1103/PhysRevLett.107.130402} {\bibfield  {journal} {\bibinfo  {journal} {Phys. Rev. Lett.}\ }\textbf {\bibinfo {volume} {107}},\ \bibinfo {pages} {130402} (\bibinfo {year} {2011})}\BibitemShut {NoStop}%
\bibitem [{\citenamefont {Knee}\ \emph {et~al.}(2012)\citenamefont {Knee}, \citenamefont {Simmons}, \citenamefont {Gauger}, \citenamefont {Morton}, \citenamefont {Riemann}, \citenamefont {Abrosimov}, \citenamefont {Becker}, \citenamefont {Pohl}, \citenamefont {Itoh}, \citenamefont {Thewalt} \emph {et~al.}}]{knee2012violation}%
  \BibitemOpen
  \bibfield  {author} {\bibinfo {author} {\bibfnamefont {G.~C.}\ \bibnamefont {Knee}}, \bibinfo {author} {\bibfnamefont {S.}~\bibnamefont {Simmons}}, \bibinfo {author} {\bibfnamefont {E.~K.}\ \bibnamefont {Gauger}}, \bibinfo {author} {\bibfnamefont {J.~J.~L.}\ \bibnamefont {Morton}}, \bibinfo {author} {\bibfnamefont {H.}~\bibnamefont {Riemann}}, \bibinfo {author} {\bibfnamefont {N.~V.}\ \bibnamefont {Abrosimov}}, \bibinfo {author} {\bibfnamefont {P.}~\bibnamefont {Becker}}, \bibinfo {author} {\bibfnamefont {H.-J.}\ \bibnamefont {Pohl}}, \bibinfo {author} {\bibfnamefont {K.~M.}\ \bibnamefont {Itoh}}, \bibinfo {author} {\bibfnamefont {M.~L.~W.}\ \bibnamefont {Thewalt}}, \emph {et~al.},\ }\bibfield  {title} {\bibinfo {title} {Violation of a leggett--garg inequality with ideal non-invasive measurements},\ }\href {https://doi.org/10.1038/ncomms1614} {\bibfield  {journal} {\bibinfo  {journal} {Nature Communications}\ }\textbf {\bibinfo {volume} {3}},\ \bibinfo {pages} {606} (\bibinfo {year} {2012})}\BibitemShut
  {NoStop}%
\bibitem [{\citenamefont {Huffman}\ and\ \citenamefont {Mizel}(2017)}]{LGIQC1}%
  \BibitemOpen
  \bibfield  {author} {\bibinfo {author} {\bibfnamefont {E.}~\bibnamefont {Huffman}}\ and\ \bibinfo {author} {\bibfnamefont {A.}~\bibnamefont {Mizel}},\ }\bibfield  {title} {\bibinfo {title} {Violation of noninvasive macrorealism by a superconducting qubit: Implementation of a leggett-garg test that addresses the clumsiness loophole},\ }\href {https://doi.org/10.1103/PhysRevA.95.032131} {\bibfield  {journal} {\bibinfo  {journal} {Phys. Rev. A}\ }\textbf {\bibinfo {volume} {95}},\ \bibinfo {pages} {032131} (\bibinfo {year} {2017})}\BibitemShut {NoStop}%
\bibitem [{\citenamefont {Ku}\ \emph {et~al.}(2020)\citenamefont {Ku}, \citenamefont {Lambert}, \citenamefont {Chan}, \citenamefont {Emary}, \citenamefont {Chen},\ and\ \citenamefont {Nori}}]{LGIQC2}%
  \BibitemOpen
  \bibfield  {author} {\bibinfo {author} {\bibfnamefont {H.-Y.}\ \bibnamefont {Ku}}, \bibinfo {author} {\bibfnamefont {N.}~\bibnamefont {Lambert}}, \bibinfo {author} {\bibfnamefont {F.-J.}\ \bibnamefont {Chan}}, \bibinfo {author} {\bibfnamefont {C.}~\bibnamefont {Emary}}, \bibinfo {author} {\bibfnamefont {Y.-N.}\ \bibnamefont {Chen}},\ and\ \bibinfo {author} {\bibfnamefont {F.}~\bibnamefont {Nori}},\ }\bibfield  {title} {\bibinfo {title} {Experimental test of non-macrorealistic cat states in the cloud},\ }\href {https://doi.org/10.1038/s41534-020-00321-x} {\bibfield  {journal} {\bibinfo  {journal} {npj Quantum Inf}\ }\textbf {\bibinfo {volume} {6}},\ \bibinfo {pages} {98} (\bibinfo {year} {2020})}\BibitemShut {NoStop}%
\bibitem [{\citenamefont {Santini}\ and\ \citenamefont {Vitale}(2022)}]{LGIQC3}%
  \BibitemOpen
  \bibfield  {author} {\bibinfo {author} {\bibfnamefont {A.}~\bibnamefont {Santini}}\ and\ \bibinfo {author} {\bibfnamefont {V.}~\bibnamefont {Vitale}},\ }\bibfield  {title} {\bibinfo {title} {Experimental violations of leggett-garg inequalities on a quantum computer},\ }\href {https://doi.org/10.1103/PhysRevA.105.032610} {\bibfield  {journal} {\bibinfo  {journal} {Phys. Rev. A}\ }\textbf {\bibinfo {volume} {105}},\ \bibinfo {pages} {032610} (\bibinfo {year} {2022})}\BibitemShut {NoStop}%
\bibitem [{\citenamefont {Nath}\ \emph {et~al.}(2025)\citenamefont {Nath}, \citenamefont {Sinha},\ and\ \citenamefont {Sinha}}]{LGIQC4}%
  \BibitemOpen
  \bibfield  {author} {\bibinfo {author} {\bibfnamefont {P.~P.}\ \bibnamefont {Nath}}, \bibinfo {author} {\bibfnamefont {A.}~\bibnamefont {Sinha}},\ and\ \bibinfo {author} {\bibfnamefont {U.}~\bibnamefont {Sinha}},\ }\bibfield  {title} {\bibinfo {title} {Certified random number generation using quantum computers},\ }\bibfield  {journal} {\bibinfo  {journal} {Frontiers in Quantum Science and Technology}\ }\textbf {\bibinfo {volume} {Volume 4 - 2025}},\ \href {https://doi.org/10.3389/frqst.2025.1661544} {10.3389/frqst.2025.1661544} (\bibinfo {year} {2025})\BibitemShut {NoStop}%
\bibitem [{\citenamefont {Melegari}\ \emph {et~al.}(2024)\citenamefont {Melegari}, \citenamefont {Cardi},\ and\ \citenamefont {Solinas}}]{LGIQC5}%
  \BibitemOpen
  \bibfield  {author} {\bibinfo {author} {\bibfnamefont {D.}~\bibnamefont {Melegari}}, \bibinfo {author} {\bibfnamefont {M.}~\bibnamefont {Cardi}},\ and\ \bibinfo {author} {\bibfnamefont {P.}~\bibnamefont {Solinas}},\ }\href {https://doi.org/10.48550/arXiv.2502.17040} {\bibinfo {title} {Quantum simulations of macrorealism violation via the qndm protocol}} (\bibinfo {year} {2024}),\ \Eprint {https://arxiv.org/abs/2502.17040} {arXiv:2502.17040 [quant-ph]} \BibitemShut {NoStop}%
\bibitem [{\citenamefont {Kofler}\ and\ \citenamefont {Brukner}(2007)}]{large_spin1}%
  \BibitemOpen
  \bibfield  {author} {\bibinfo {author} {\bibfnamefont {J.}~\bibnamefont {Kofler}}\ and\ \bibinfo {author} {\bibfnamefont {C.}~\bibnamefont {Brukner}},\ }\bibfield  {title} {\bibinfo {title} {Classical world arising out of quantum physics under the restriction of coarse-grained measurements},\ }\href {https://doi.org/10.1103/PhysRevLett.99.180403} {\bibfield  {journal} {\bibinfo  {journal} {Phys. Rev. Lett.}\ }\textbf {\bibinfo {volume} {99}},\ \bibinfo {pages} {180403} (\bibinfo {year} {2007})}\BibitemShut {NoStop}%
\bibitem [{\citenamefont {Schlosshauer}(2007)}]{schlosshauer_quantum_2007}%
  \BibitemOpen
  \bibfield  {author} {\bibinfo {author} {\bibfnamefont {M.}~\bibnamefont {Schlosshauer}},\ }\href@noop {} {\emph {\bibinfo {title} {{The quantum-to-classical transition and decoherence}}}}\ (\bibinfo  {publisher} {Springer International Publishing},\ \bibinfo {year} {2007})\ \bibinfo {note} {\href{https://doi.org/10.1007/978-3-540-35775-9}{10.1007/978-3-540-35775-9}}\BibitemShut {NoStop}%
\bibitem [{\citenamefont {Das}\ \emph {et~al.}(2024)\citenamefont {Das}, \citenamefont {Home}, \citenamefont {Ulbricht},\ and\ \citenamefont {Bose}}]{MassMR2024}%
  \BibitemOpen
  \bibfield  {author} {\bibinfo {author} {\bibfnamefont {D.}~\bibnamefont {Das}}, \bibinfo {author} {\bibfnamefont {D.}~\bibnamefont {Home}}, \bibinfo {author} {\bibfnamefont {H.}~\bibnamefont {Ulbricht}},\ and\ \bibinfo {author} {\bibfnamefont {S.}~\bibnamefont {Bose}},\ }\bibfield  {title} {\bibinfo {title} {Mass-independent scheme to test the quantumness of a massive object},\ }\href {https://doi.org/10.1103/PhysRevLett.132.030202} {\bibfield  {journal} {\bibinfo  {journal} {Phys. Rev. Lett.}\ }\textbf {\bibinfo {volume} {132}},\ \bibinfo {pages} {030202} (\bibinfo {year} {2024})}\BibitemShut {NoStop}%
\bibitem [{\citenamefont {Bibak}\ \emph {et~al.}(2025)\citenamefont {Bibak}, \citenamefont {Cepollaro} \emph {et~al.}}]{classicality_emergence}%
  \BibitemOpen
  \bibfield  {author} {\bibinfo {author} {\bibfnamefont {F.}~\bibnamefont {Bibak}}, \bibinfo {author} {\bibfnamefont {C.}~\bibnamefont {Cepollaro}}, \emph {et~al.},\ }\bibfield  {title} {\bibinfo {title} {The classical limit of quantum mechanics through coarse-grained measurements},\ }\bibfield  {journal} {\bibinfo  {journal} {arXiv}\ }\href {https://doi.org/quant-ph/2503.15642} {quant-ph/2503.15642} (\bibinfo {year} {2025})\BibitemShut {NoStop}%
\bibitem [{\citenamefont {Braccini}\ \emph {et~al.}()\citenamefont {Braccini}, \citenamefont {Das}, \citenamefont {Zindorf}, \citenamefont {Hogan}, \citenamefont {Morton},\ and\ \citenamefont {Bose}}]{LargeSpinParallel}%
  \BibitemOpen
  \bibfield  {author} {\bibinfo {author} {\bibfnamefont {L.}~\bibnamefont {Braccini}}, \bibinfo {author} {\bibfnamefont {D.}~\bibnamefont {Das}}, \bibinfo {author} {\bibfnamefont {B.}~\bibnamefont {Zindorf}}, \bibinfo {author} {\bibfnamefont {S.~D.}\ \bibnamefont {Hogan}}, \bibinfo {author} {\bibfnamefont {J.~J.~L.}\ \bibnamefont {Morton}},\ and\ \bibinfo {author} {\bibfnamefont {S.}~\bibnamefont {Bose}},\ }\bibfield  {title} {\bibinfo {title} {Collective nonclassicality of a macroscopic qubit-ensemble via measurement induced disturbance},\ }\href@noop {} {\bibinfo  {journal} {To Appear Parallelly}\ }\BibitemShut {NoStop}%
\bibitem [{\citenamefont {Joarder}\ \emph {et~al.}(2022)\citenamefont {Joarder}, \citenamefont {Saha}, \citenamefont {Home},\ and\ \citenamefont {Sinha}}]{Joarder_Loophole_2022}%
  \BibitemOpen
\bibfield  {journal} {  }\bibfield  {author} {\bibinfo {author} {\bibfnamefont {K.}~\bibnamefont {Joarder}}, \bibinfo {author} {\bibfnamefont {D.}~\bibnamefont {Saha}}, \bibinfo {author} {\bibfnamefont {D.}~\bibnamefont {Home}},\ and\ \bibinfo {author} {\bibfnamefont {U.}~\bibnamefont {Sinha}},\ }\bibfield  {title} {\bibinfo {title} {Loophole-free interferometric test of macrorealism using heralded single photons},\ }\bibfield  {journal} {\bibinfo  {journal} {PRX Quantum}\ }\textbf {\bibinfo {volume} {3}},\ \href {https://doi.org/10.1103/PRXQuantum.3.010307} {10.1103/PRXQuantum.3.010307} (\bibinfo {year} {2022})\BibitemShut {NoStop}%
\bibitem [{\citenamefont {Wilde}\ and\ \citenamefont {Mizel}(2012)}]{Wilde_Clumsiness_2012}%
  \BibitemOpen
  \bibfield  {author} {\bibinfo {author} {\bibfnamefont {M.~M.}\ \bibnamefont {Wilde}}\ and\ \bibinfo {author} {\bibfnamefont {A.}~\bibnamefont {Mizel}},\ }\bibfield  {title} {\bibinfo {title} {Addressing the clumsiness loophole in a leggett-garg test of macrorealism},\ }\bibfield  {journal} {\bibinfo  {journal} {Foundations of Physics}\ }\textbf {\bibinfo {volume} {42}},\ \href {https://doi.org/10.1007/s10701-011-9598-4} {10.1007/s10701-011-9598-4} (\bibinfo {year} {2012})\BibitemShut {NoStop}%
\bibitem [{\citenamefont {Knee}\ \emph {et~al.}(2016{\natexlab{b}})\citenamefont {Knee}, \citenamefont {Kakuyanagi}, \citenamefont {Yeh}, \citenamefont {Matsuzaki}, \citenamefont {Toida}, \citenamefont {Yamaguchi}, \citenamefont {Saito}, \citenamefont {Leggett},\ and\ \citenamefont {Munro}}]{Knee_strict_2016}%
  \BibitemOpen
  \bibfield  {author} {\bibinfo {author} {\bibfnamefont {G.~C.}\ \bibnamefont {Knee}}, \bibinfo {author} {\bibfnamefont {K.}~\bibnamefont {Kakuyanagi}}, \bibinfo {author} {\bibfnamefont {M.-C.}\ \bibnamefont {Yeh}}, \bibinfo {author} {\bibfnamefont {Y.}~\bibnamefont {Matsuzaki}}, \bibinfo {author} {\bibfnamefont {H.}~\bibnamefont {Toida}}, \bibinfo {author} {\bibfnamefont {H.}~\bibnamefont {Yamaguchi}}, \bibinfo {author} {\bibfnamefont {S.}~\bibnamefont {Saito}}, \bibinfo {author} {\bibfnamefont {A.~J.}\ \bibnamefont {Leggett}},\ and\ \bibinfo {author} {\bibfnamefont {W.~J.}\ \bibnamefont {Munro}},\ }\bibfield  {title} {\bibinfo {title} {A strict experimental test of macroscopic realism in a superconducting flux qubit},\ }\bibfield  {journal} {\bibinfo  {journal} {Nature Communications}\ }\textbf {\bibinfo {volume} {7}},\ \href {https://doi.org/10.1038/ncomms13253} {10.1038/ncomms13253} (\bibinfo {year} {2016}{\natexlab{b}})\BibitemShut {NoStop}%
\bibitem [{\citenamefont {Lambert}\ \emph {et~al.}(2020)\citenamefont {Lambert}, \citenamefont {Chan}, \citenamefont {Emary}, \citenamefont {Chen},\ and\ \citenamefont {Nori}}]{Lambert_Experimental_2020}%
  \BibitemOpen
  \bibfield  {author} {\bibinfo {author} {\bibfnamefont {N.}~\bibnamefont {Lambert}}, \bibinfo {author} {\bibfnamefont {F.-J.}\ \bibnamefont {Chan}}, \bibinfo {author} {\bibfnamefont {C.}~\bibnamefont {Emary}}, \bibinfo {author} {\bibfnamefont {Y.-N.}\ \bibnamefont {Chen}},\ and\ \bibinfo {author} {\bibfnamefont {F.}~\bibnamefont {Nori}},\ }\bibfield  {title} {\bibinfo {title} {Experimental test of non-macrorealistic cat states in the cloud},\ }\bibfield  {journal} {\bibinfo  {journal} {npj Quantum Information}\ }\textbf {\bibinfo {volume} {6}},\ \href {https://doi.org/10.1038/s41534-020-00321-x} {10.1038/s41534-020-00321-x} (\bibinfo {year} {2020})\BibitemShut {NoStop}%
\bibitem [{\citenamefont {Kutin}\ \emph {et~al.}(2007)\citenamefont {Kutin}, \citenamefont {Moulton},\ and\ \citenamefont {Smithline}}]{kutin2007computation}%
  \BibitemOpen
  \bibfield  {author} {\bibinfo {author} {\bibfnamefont {S.~A.}\ \bibnamefont {Kutin}}, \bibinfo {author} {\bibfnamefont {D.~P.}\ \bibnamefont {Moulton}},\ and\ \bibinfo {author} {\bibfnamefont {L.~M.}\ \bibnamefont {Smithline}},\ }\href {https://arxiv.org/abs/quant-ph/0701194} {\bibinfo {title} {Computation at a distance}} (\bibinfo {year} {2007}),\ \Eprint {https://arxiv.org/abs/quant-ph/0701194} {arXiv:quant-ph/0701194 [quant-ph]} \BibitemShut {NoStop}%
\bibitem [{\citenamefont {IBM}(2025)}]{IBM}%
  \BibitemOpen
  \bibfield  {author} {\bibinfo {author} {\bibnamefont {IBM}},\ }\href@noop {} {\bibinfo {title} {Quantum}} (\bibinfo {year} {2025}),\ \bibinfo {note} {\href{https://quantum.cloud.ibm.com/}{https://quantum.cloud.ibm.com/}}\BibitemShut {NoStop}%
\bibitem [{\citenamefont {Antonio}\ \emph {et~al.}(2015)\citenamefont {Antonio}, \citenamefont {Randall}, \citenamefont {Hensinger}, \citenamefont {Morley},\ and\ \citenamefont {Bose}}]{antonio2015classicalcomputationquantumbits}%
  \BibitemOpen
  \bibfield  {author} {\bibinfo {author} {\bibfnamefont {B.}~\bibnamefont {Antonio}}, \bibinfo {author} {\bibfnamefont {J.}~\bibnamefont {Randall}}, \bibinfo {author} {\bibfnamefont {W.~K.}\ \bibnamefont {Hensinger}}, \bibinfo {author} {\bibfnamefont {G.~W.}\ \bibnamefont {Morley}},\ and\ \bibinfo {author} {\bibfnamefont {S.}~\bibnamefont {Bose}},\ }\href {https://arxiv.org/abs/1509.03420} {\bibinfo {title} {Classical computation by quantum bits}} (\bibinfo {year} {2015}),\ \Eprint {https://arxiv.org/abs/1509.03420} {arXiv:1509.03420 [quant-ph]} \BibitemShut {NoStop}%
\end{thebibliography}

%

\onecolumngrid

\appendix

\section{Quantum Circuits Optimization: H-method}
\label{app:quantum_circ} 

\qc{og_parity} provides the H-method shown in 
Fig.~\ref{fig:circuit_1}.(a) for $N=4$, decomposed using the definitions in Fig.~\ref{fig:circuit_1}.(c).
As can be seen, the resulting circuit is comprised of long range CNOT gates which cannot be implemented directly in some architectures.
We wish to find an equivalent circuit which can be implemented efficiently under the restricted connectivity of IBM's superconducting qubits.
While the quantum computers provided by IBM allow heavy-hex connectivity, we focus on linear-nearest-neighbor (LNN) connectivity which is more restrictive in most cases, and can be efficiently mapped to heavy-hex connectivity.
We show that \qc{og_parity} is equivalent to 
\qc{full_final_depth_parity}, and that in case the result of the measurement applied on the $a_1$ ancilla can be discarded,
\qc{final_depth_parity} can be used instead. 
\[
\scalebox{0.8}{
\Qcircuit @C=1.0em @R=0.2em @!R { 
\nghost{\mathrm{c_p :  }} & \lstick{\mathrm{c_p :  }} &  \controlo \cw \ar @{-} [1,0] & \cw & \cw & \cw & \cw & \cw & \cw & \cw & \cw & \cw & \cw & \cw & \cw & \cw & \cw \\
	 	\nghost{{a}_{1} :  } & \lstick{{a}_{1} :  } & \gate{\mathrm{H}}  & \targ & \targ & \targ & \targ & \meter \barrier[0em]{5} & \qw & \qw & \qw & \qw & \qw & \qw & \qw & \qw & \qw\\
	 	\nghost{{q}_{1} :  } & \lstick{{q}_{1} :  } & \gate{\mathrm{\theta}} & \ctrl{-1} & \qw & \qw & \qw & \qw & \qw & \gate{\mathrm{\theta}} & \ctrl{4} & \qw & \qw & \qw & \qw & \qw & \qw\\
	 	\nghost{{q}_{2} :  } & \lstick{{q}_{2} :  } & \gate{\mathrm{\theta}} & \qw & \ctrl{-2} & \qw & \qw & \qw & \qw & \gate{\mathrm{\theta}} & \qw & \ctrl{3} & \qw & \qw & \qw & \qw & \qw\\
	 	\nghost{{q}_{3} :  } & \lstick{{q}_{3} :  } & \gate{\mathrm{\theta}} & \qw & \qw & \ctrl{-3} & \qw & \qw & \qw & \gate{\mathrm{\theta}} & \qw & \qw & \ctrl{2} & \qw & \qw & \qw & \qw\\
	 	\nghost{{q}_{4} :  } & \lstick{{q}_{4} :  } & \gate{\mathrm{\theta}} & \qw & \qw & \qw & \ctrl{-4} & \qw & \qw & \gate{\mathrm{\theta}} & \qw & \qw & \qw & \ctrl{1} & \qw & \qw & \qw\\
	 	\nghost{{a}_{2} :  } & \lstick{{a}_{2} :  } & \qw & \qw & \qw & \qw & \qw & \qw & \qw & \qw & \targ & \targ & \targ & \targ & \meter & \qw & \qw\\
	 	\nghost{\mathrm{{c_{1,2}} :  }} & \lstick{\mathrm{{c_{1,2}} :  }} & \lstick{/_{_{2}}} \cw & \cw & \cw & \cw & \cw & \dstick{_{_{\hspace{0.0em}{c_1}}}} \cw \ar @{<=} [-6,0] & \cw & \cw & \cw & \cw & \cw & \cw & \dstick{_{_{\hspace{0.0em}{c_2}}}} \cw \ar @{<=} [-1,0] & \cw & \cw\\
 }}
 \qcref{og_parity}
\]

The measurement on the $a_1$ ancilla can be commuted to the end of the circuit. In addition, we apply two SWAP chains, and adjust the qubit permutation between these chains accordingly, in order to obtain \qc{og_swaps_parity}, which is equivalent to \qc{og_parity}.

\[
\scalebox{0.7}{
\Qcircuit @C=1.0em @R=0.2em @!R { 
\nghost{\mathrm{c_p :  }} & \lstick{\mathrm{c_p :  }} &  \controlo \cw \ar @{-} [1,0]  & \cw & \cw & \cw & \cw & \cw & \cw & \cw & \cw & \cw & \cw & \cw & \cw & \cw & \cw & \cw & \cw & \cw & \cw & \cw & \cw & \cw & \cw & \cw & \cw\\
	 	\nghost{{a}_{1} :  } & \lstick{{a}_{1} :  } & \gate{\mathrm{H}} & \targ & \targ & \targ & \targ & \qswap & \qw & \qw & \qw \barrier[0em]{5} & \qw & \gate{\mathrm{\theta}} & \ctrl{5} & \qw & \qw & \qw \barrier[0em]{5} & \qw & \qw & \qw & \qw & \qswap \barrier[0em]{5} & \qw & \meter & \qw & \qw & \qw\\
	 	\nghost{{q}_{1} :  } & \lstick{{q}_{1} :  } & \gate{\mathrm{\theta}} & \ctrl{-1} & \qw & \qw & \qw & \qswap \qwx[-1] & \qswap & \qw & \qw & \qw & \gate{\mathrm{\theta}} & \qw & \ctrl{4} & \qw & \qw & \qw & \qw & \qw & \qswap & \qswap \qwx[-1] & \qw & \qw & \qw & \qw & \qw\\
	 	\nghost{{q}_{2} :  } & \lstick{{q}_{2} :  } & \gate{\mathrm{\theta}} & \qw & \ctrl{-2} & \qw & \qw & \qw & \qswap \qwx[-1] & \qswap & \qw & \qw & \gate{\mathrm{\theta}} & \qw & \qw & \ctrl{3} & \qw & \qw & \qw & \qswap & \qswap \qwx[-1] & \qw & \qw & \qw & \qw & \qw & \qw\\
	 	\nghost{{q}_{3} :  } & \lstick{{q}_{3} :  } & \gate{\mathrm{\theta}} & \qw & \qw & \ctrl{-3} & \qw & \qw & \qw & \qswap \qwx[-1] & \qswap & \qw & \gate{\mathrm{\theta}} & \qw & \qw & \qw & \ctrl{2} & \qw & \qswap & \qswap \qwx[-1] & \qw & \qw & \qw & \qw & \qw & \qw & \qw\\
	 	\nghost{{q}_{4} :  } & \lstick{{q}_{4} :  } & \gate{\mathrm{\theta}} & \qw & \qw & \qw & \ctrl{-4} & \qw & \qw & \qw & \qswap \qwx[-1] & \qw & \qw & \qw & \qw & \qw & \qw & \qw & \qswap \qwx[-1] & \qw & \qw & \qw & \qw & \qw & \qw & \qw & \qw\\
	 	\nghost{{a}_{2} :  } & \lstick{{a}_{2} :  } & \qw & \qw & \qw & \qw & \qw & \qw & \qw & \qw & \qw & \qw & \qw & \targ & \targ & \targ & \targ & \qw & \qw & \qw & \qw & \qw & \qw & \qw & \meter & \qw & \qw\\
	 	\nghost{\mathrm{{c_{1,2}} :  }} & \lstick{\mathrm{{c_{1,2}} :  }} & \lstick{/_{_{2}}} \cw & \cw & \cw & \cw & \cw & \cw & \cw & \cw & \cw & \cw & \cw & \cw & \cw & \cw & \cw & \cw & \cw & \cw & \cw & \cw & \cw & \dstick{_{_{\hspace{0.0em}{c_1}}}} \cw \ar @{<=} [-6,0] & \dstick{_{_{\hspace{0.0em}{c_2}}}} \cw \ar @{<=} [-1,0] & \cw & \cw\\
 }}
\qcref{og_swaps_parity}
\]
Next, the CNOT gates on the left of the first barrier (vertical dashed line) in \qc{og_swaps_parity} can commute with SWAP gates on their right, while changing the CNOT target qubit at each such commutation. This produces the structure on the left of the first barrier in \qc{second_swaps_parity}. The remaining long-range CNOT gates can be converted to an efficient LNN circuit by repeatedly applying the well-known identity:
\[
\scalebox{1.0}{
 \Qcircuit @C=1.0em @R=0.8em @!R { 
	 	\nghost{} & \lstick{} & \ctrl{2} & \qw & \qw & \qw\\
	 	\nghost{} & \lstick{} & \qw & \ctrl{1} & \qw & \qw\\
	 	\nghost{} & \lstick{} & \targ & \targ & \qw & \qw\\
 } \hspace{5mm}\raisebox{-7mm}{=}\hspace{0mm}
\Qcircuit @C=1.0em @R=0.8em @!R { 
	 	\nghost{} & \lstick{} & \ctrl{1} & \qw & \ctrl{1} & \qw & \qw\\
	 	\nghost{} & \lstick{} & \targ & \ctrl{1} & \targ & \qw & \qw\\
	 	\nghost{} & \lstick{} & \qw & \targ & \qw & \qw & \qw\\
 }
 }
 \qcref{identity_cir}
\]
Theses steps produce \qc{second_swaps_parity} in which all two-qubit gates are applied on nearest neighbouring qubits.
\[
\scalebox{0.6}{
\Qcircuit @C=1.0em @R=0.2em @!R {
\nghost{\mathrm{c_p :  }} & \lstick{\mathrm{c_p :  }} &  \controlo \cw \ar @{-} [1,0]  & \cw & \cw & \cw & \cw & \cw & \cw & \cw & \cw & \cw & \cw & \cw & \cw & \cw & \cw & \cw & \cw & \cw & \cw & \cw & \cw & \cw & \cw & \cw & \cw & \cw & \cw & \cw & \cw & \cw \\
	 	\nghost{{a}_{1} :  } & \lstick{{a}_{1} :  } & \gate{\mathrm{H}} & \targ & \qswap & \qw & \qw & \qw & \qw & \qw & \qw \barrier[0em]{5} & \qw & \gate{\mathrm{\theta}} & \ctrl{1} & \qw & \qw & \qw & \qw & \qw & \qw & \qw & \ctrl{1} \barrier[0em]{5} & \qw & \qw & \qw & \qw & \qswap \barrier[0em]{5} & \qw & \meter & \qw & \qw & \qw\\
	 	\nghost{{q}_{1} :  } & \lstick{{q}_{1} :  } & \gate{\mathrm{\theta}} & \ctrl{-1} & \qswap \qwx[-1] & \targ & \qswap & \qw & \qw & \qw & \qw & \qw & \gate{\mathrm{\theta}} & \targ & \ctrl{1} & \qw & \qw & \qw & \qw & \qw & \ctrl{1} & \targ & \qw & \qw & \qw & \qswap & \qswap \qwx[-1] & \qw & \qw & \qw & \qw & \qw\\
	 	\nghost{{q}_{2} :  } & \lstick{{q}_{2} :  } & \gate{\mathrm{\theta}} & \qw & \qw & \ctrl{-1} & \qswap \qwx[-1] & \targ & \qswap & \qw & \qw & \qw & \gate{\mathrm{\theta}} & \qw & \targ & \ctrl{1} & \qw & \qw & \qw & \ctrl{1} & \targ & \qw & \qw & \qw & \qswap & \qswap \qwx[-1] & \qw & \qw & \qw & \qw & \qw & \qw\\
	 	\nghost{{q}_{3} :  } & \lstick{{q}_{3} :  } & \gate{\mathrm{\theta}} & \qw & \qw & \qw & \qw & \ctrl{-1} & \qswap \qwx[-1] & \targ & \qswap & \qw & \gate{\mathrm{\theta}} & \qw & \qw & \targ & \qswap & \qw & \qswap & \targ & \qw & \qw & \qw & \qswap & \qswap \qwx[-1] & \qw & \qw & \qw & \qw & \qw & \qw & \qw\\
	 	\nghost{{q}_{4} :  } & \lstick{{q}_{4} :  } & \gate{\mathrm{\theta}} & \qw & \qw & \qw & \qw & \qw & \qw & \ctrl{-1} & \qswap \qwx[-1] & \qw & \qw & \qw & \qw & \qw & \qswap \qwx[-1] & \ctrl{1} & \qswap \qwx[-1] & \qw & \qw & \qw & \qw & \qswap \qwx[-1] & \qw & \qw & \qw & \qw & \qw & \qw & \qw & \qw\\
	 	\nghost{{a}_{2} :  } & \lstick{{a}_{2} :  } & \qw & \qw & \qw & \qw & \qw & \qw & \qw & \qw & \qw & \qw & \qw & \qw & \qw & \qw & \qw & \targ & \qw & \qw & \qw & \qw & \qw & \qw & \qw & \qw & \qw & \qw & \qw & \meter & \qw & \qw\\
	 	\nghost{\mathrm{{c_{1,2}} :  }} & \lstick{\mathrm{{c_{1,2}} :  }} & \lstick{/_{_{2}}} \cw & \cw & \cw & \cw & \cw & \cw & \cw & \cw & \cw & \cw & \cw & \cw & \cw & \cw & \cw & \cw & \cw & \cw & \cw & \cw & \cw & \cw & \cw & \cw & \cw & \cw & \dstick{_{_{\hspace{0.0em}{c_1}}}} \cw \ar @{<=} [-6,0] & \dstick{_{_{\hspace{0.0em}{c_2}}}} \cw \ar @{<=} [-1,0] & \cw & \cw\\
 }}
 \qcref{second_swaps_parity}
\]
\qc{full_final_depth_parity} is achieved by decomposing SWAP gates as three CNOT gates, applying cancellations and commuting CNOT gates which are controlled by the same qubit. Moreover, the measurement on $a_2$ is commuted to the left.

\[
\scalebox{0.6}{
\Qcircuit @C=1.0em @R=0.2em @!R { 
\nghost{\mathrm{c_p :  }} & \lstick{\mathrm{c_p :  }} &  \controlo \cw \ar @{-} [1,0]  & \cw & \cw & \cw & \cw & \cw & \cw & \cw & \cw & \cw & \cw & \cw & \cw & \cw & \cw & \cw & \cw & \cw & \cw & \cw & \cw & \cw & \cw & \cw & \cw & \cw & \cw\\
	 	\nghost{{a}_{1} :  } & \lstick{{a}_{1} :  } & \gate{\mathrm{H}} & \ctrl{1} & \qw & \targ & \qw & \qw & \qw \barrier[0em]{5} & \qw & \gate{\mathrm{\theta}} & \ctrl{1} & \qw & \qw & \qw & \qw & \qw & \qw \barrier[0em]{5} & \qw & \qw & \qw & \qw & \qw & \ctrl{1} & \qw & \qswap & \meter & \qw & \qw\\
	 	\nghost{{q}_{1} :  } & \lstick{{q}_{1} :  } & \gate{\mathrm{\theta}} & \targ & \ctrl{1} & \ctrl{-1} & \targ & \qw & \qw & \qw & \gate{\mathrm{\theta}} & \targ & \ctrl{1} & \qw & \qw & \qw & \qw & \qw & \qw & \qw & \qw & \qw & \ctrl{1} & \targ & \qswap & \qswap \qwx[-1] & \qw & \qw & \qw\\
	 	\nghost{{q}_{2} :  } & \lstick{{q}_{2} :  } & \gate{\mathrm{\theta}} & \qw & \targ & \ctrl{1} & \ctrl{-1} & \targ & \qw & \qw & \gate{\mathrm{\theta}} & \qw & \targ & \ctrl{1} & \qw & \qw & \qw & \qw & \qw & \qw & \qw & \ctrl{1} & \targ & \qswap & \qswap \qwx[-1] & \qw & \qw & \qw & \qw\\
	 	\nghost{{q}_{3} :  } & \lstick{{q}_{3} :  } & \gate{\mathrm{\theta}} & \qw & \qw & \targ & \ctrl{1} & \ctrl{-1} & \targ & \qw & \gate{\mathrm{\theta}} & \qw & \qw & \targ & \targ & \ctrl{1} & \qw & \qw & \qw & \targ & \qswap & \targ & \qswap & \qswap \qwx[-1] & \qw & \qw & \qw & \qw & \qw\\
	 	\nghost{{q}_{4} :  } & \lstick{{q}_{4} :  } & \gate{\mathrm{\theta}} & \qw & \qw & \qw & \targ & \qw & \ctrl{-1} & \qw & \qw & \qw & \qw & \qw & \ctrl{-1} & \targ & \ctrl{1} & \qw & \qw & \ctrl{-1} & \qswap \qwx[-1] & \qw & \qswap \qwx[-1] & \qw & \qw & \qw & \qw & \qw & \qw\\
	 	\nghost{{a}_{2} :  } & \lstick{{a}_{2} :  } & \qw & \qw & \qw & \qw & \qw & \qw & \qw & \qw & \qw & \qw & \qw & \qw & \qw & \qw & \targ & \meter & \qw & \qw & \qw & \qw & \qw & \qw & \qw & \qw & \qw & \qw & \qw\\
	 	\nghost{\mathrm{{c_{1,2}} :  }} & \lstick{\mathrm{{c_{1,2}} :  }} & \lstick{/_{_{2}}} \cw & \cw & \cw & \cw & \cw & \cw & \cw & \cw & \cw & \cw & \cw & \cw & \cw & \cw & \cw & \dstick{_{_{\hspace{0.0em}{c_2}}}} \cw \ar @{<=} [-1,0] & \cw & \cw & \cw & \cw & \cw & \cw & \cw & \cw & \dstick{_{_{\hspace{0.0em}{c_1}}}} \cw \ar @{<=} [-6,0] & \cw & \cw\\
 }
}
 \qcref{full_final_depth_parity}
\]
Following the steps above, it is clear that \qc{full_final_depth_parity} is equivalent to \qc{og_parity}.
It can be noted that the result of the measurement on $a_1$ does not effect the value of $V_{\pm}$. Moreover, once the result of the $a_2$ measurement has been obtained, it will not be changed by any gate which is applied in the circuit. Therefore, all of the gates which are applied after the second barrier in  \qc{full_final_depth_parity} can be removed. \qc{final_depth_parity} is finally obtained by removing these gates, along with the barriers, allowing additional gates to be applied in parallel.
\[
\scalebox{0.8}{
\Qcircuit @C=.5em @R=0.2em @!R {
\nghost{\mathrm{c_p :  }} & \lstick{\mathrm{c_p :  }} &  \controlo \cw \ar @{-} [1,0] & \cw & \cw & \cw & \cw & \cw & \cw & \cw & \cw & \cw & \cw & \cw & \cw & \cw & \cw  \\
	 	\nghost{{a}_{1} :  } & \lstick{{a}_{1} :  } & \gate{\mathrm{H}} & \ctrl{1} & \qw & \targ & \gate{\mathrm{\theta}} & \qw & \ctrl{1} & \qw & \qw & \qw & \qw & \qw & \qw & \qw & \qw\\
	 	\nghost{{q}_{1} :  } & \lstick{{q}_{1} :  } & \gate{\mathrm{\theta}} & \targ & \ctrl{1} & \ctrl{-1} & \targ & \gate{\mathrm{\theta}} & \targ & \ctrl{1} & \qw & \qw & \qw & \qw & \qw & \qw & \qw\\
	 	\nghost{{q}_{2} :  } & \lstick{{q}_{2} :  } & \gate{\mathrm{\theta}} & \qw & \targ & \ctrl{1} & \ctrl{-1} & \targ & \gate{\mathrm{\theta}} & \targ & \ctrl{1} & \qw & \qw & \qw & \qw & \qw & \qw\\
	 	\nghost{{q}_{3} :  } & \lstick{{q}_{3} :  } & \gate{\mathrm{\theta}} & \qw & \qw & \targ & \ctrl{1} & \ctrl{-1} & \targ & \gate{\mathrm{\theta}} & \targ & \targ & \ctrl{1} & \qw & \qw & \qw & \qw\\
	 	\nghost{{q}_{4} :  } & \lstick{{q}_{4} :  } & \gate{\mathrm{\theta}} & \qw & \qw & \qw & \targ & \qw & \ctrl{-1} & \qw & \qw & \ctrl{-1} & \targ & \ctrl{1} & \qw & \qw & \qw\\
	 	\nghost{{a}_{2} :  } & \lstick{{a}_{2} :  } & \qw & \qw & \qw & \qw & \qw & \qw & \qw & \qw & \qw & \qw & \qw & \targ & \meter & \qw & \qw\\
	 	\nghost{\mathrm{{c_{2}} :  }} & \lstick{\mathrm{{c_{2}} :  }} & \lstick{/_{_{1}}} \cw & \cw & \cw & \cw & \cw & \cw & \cw & \cw & \cw & \cw & \cw & \cw & \dstick{_{_{\hspace{0.0em}{c_2}}}} \cw \ar @{<=} [-1,0] & \cw & \cw\\
 }}
 \qcref{final_depth_parity}
\]

\qc{final_large_depth_parity} demonstrates the structure achieved by this process for a larger number of qubits. It can be noted that many gates are applied in parallel, thus reducing the depth of the circuit. This circuit over $N$ qubits is applied using $3N+O(1)$ LNN CNOT gates in depth $N+O(1)$. 

\[
\scalebox{0.55}{
\Qcircuit @C=0.2em @R=0.2em @!R { 
\nghost{\mathrm{c_p :  }} & \lstick{\mathrm{c_p :  }} &  \controlo \cw \ar @{-} [1,0]  & \cw & \cw & \cw & \cw & \cw & \cw & \cw & \cw & \cw & \cw & \cw & \cw & \cw & \cw & \cw & \cw & \cw & \cw & \cw & \cw & \cw & \cw & \cw & \cw & \cw & \cw & \cw \\
	 	\nghost{{a}_{1} :  } & \lstick{{a}_{1} :  } & \gate{\mathrm{H}} & \targ & \targ & \targ & \targ & \targ & \targ & \targ & \targ & \targ & \targ \barrier[0em]{11} & \qw & \qw & \qw & \qw & \qw & \qw & \qw & \qw & \qw & \qw & \qw & \qw \barrier[0em]{11} & \qw & \meter & \qw & \qw & \qw\\
	 	\nghost{{q}_{1} :  } & \lstick{{q}_{1} :  } & \gate{\mathrm{\theta}} & \ctrl{-1} & \qw & \qw & \qw & \qw & \qw & \qw & \qw & \qw & \qw & \qw & \gate{\mathrm{\theta}} & \ctrl{10} & \qw & \qw & \qw & \qw & \qw & \qw & \qw & \qw & \qw & \qw & \qw & \qw & \qw & \qw\\
	 	\nghost{{q}_{2} :  } & \lstick{{q}_{2} :  } & \gate{\mathrm{\theta}} & \qw & \ctrl{-2} & \qw & \qw & \qw & \qw & \qw & \qw & \qw & \qw & \qw & \gate{\mathrm{\theta}} & \qw & \ctrl{9} & \qw & \qw & \qw & \qw & \qw & \qw & \qw & \qw & \qw & \qw & \qw & \qw & \qw\\
	 	\nghost{{q}_{3} :  } & \lstick{{q}_{3} :  } & \gate{\mathrm{\theta}} & \qw & \qw & \ctrl{-3} & \qw & \qw & \qw & \qw & \qw & \qw & \qw & \qw & \gate{\mathrm{\theta}} & \qw & \qw & \ctrl{8} & \qw & \qw & \qw & \qw & \qw & \qw & \qw & \qw & \qw & \qw & \qw & \qw\\
	 	\nghost{{q}_{4} :  } & \lstick{{q}_{4} :  } & \gate{\mathrm{\theta}} & \qw & \qw & \qw & \ctrl{-4} & \qw & \qw & \qw & \qw & \qw & \qw & \qw & \gate{\mathrm{\theta}} & \qw & \qw & \qw & \ctrl{7} & \qw & \qw & \qw & \qw & \qw & \qw & \qw & \qw & \qw & \qw & \qw\\
	 	\nghost{{q}_{5} :  } & \lstick{{q}_{5} :  } & \gate{\mathrm{\theta}} & \qw & \qw & \qw & \qw & \ctrl{-5} & \qw & \qw & \qw & \qw & \qw & \qw & \gate{\mathrm{\theta}} & \qw & \qw & \qw & \qw & \ctrl{6} & \qw & \qw & \qw & \qw & \qw & \qw & \qw & \qw & \qw & \qw\\
	 	\nghost{{q}_{6} :  } & \lstick{{q}_{6} :  } & \gate{\mathrm{\theta}} & \qw & \qw & \qw & \qw & \qw & \ctrl{-6} & \qw & \qw & \qw & \qw & \qw & \gate{\mathrm{\theta}} & \qw & \qw & \qw & \qw & \qw & \ctrl{5} & \qw & \qw & \qw & \qw & \qw & \qw & \qw & \qw & \qw\\
	 	\nghost{{q}_{7} :  } & \lstick{{q}_{7} :  } & \gate{\mathrm{\theta}} & \qw & \qw & \qw & \qw & \qw & \qw & \ctrl{-7} & \qw & \qw & \qw & \qw & \gate{\mathrm{\theta}} & \qw & \qw & \qw & \qw & \qw & \qw & \ctrl{4} & \qw & \qw & \qw & \qw & \qw & \qw & \qw & \qw\\
	 	\nghost{{q}_{8} :  } & \lstick{{q}_{8} :  } & \gate{\mathrm{\theta}} & \qw & \qw & \qw & \qw & \qw & \qw & \qw & \ctrl{-8} & \qw & \qw & \qw & \gate{\mathrm{\theta}} & \qw & \qw & \qw & \qw & \qw & \qw & \qw & \ctrl{3} & \qw & \qw & \qw & \qw & \qw & \qw & \qw\\
	 	\nghost{{q}_{9} :  } & \lstick{{q}_{9} :  } & \gate{\mathrm{\theta}} & \qw & \qw & \qw & \qw & \qw & \qw & \qw & \qw & \ctrl{-9} & \qw & \qw & \gate{\mathrm{\theta}} & \qw & \qw & \qw & \qw & \qw & \qw & \qw & \qw & \ctrl{2} & \qw & \qw & \qw & \qw & \qw & \qw\\
	 	\nghost{{q}_{10} :  } & \lstick{{q}_{10} :  } & \gate{\mathrm{\theta}} & \qw & \qw & \qw & \qw & \qw & \qw & \qw & \qw & \qw & \ctrl{-10} & \qw & \gate{\mathrm{\theta}} & \qw & \qw & \qw & \qw & \qw & \qw & \qw & \qw & \qw & \ctrl{1} & \qw & \qw & \qw & \qw & \qw\\
	 	\nghost{{a}_{2} :  } & \lstick{{a}_{2} :  } & \qw & \qw & \qw & \qw & \qw & \qw & \qw & \qw & \qw & \qw & \qw & \qw & \qw & \targ & \targ & \targ & \targ & \targ & \targ & \targ & \targ & \targ & \targ & \qw & \qw & \meter & \qw & \qw\\
	 	\nghost{\mathrm{{c_{1,2}} :  }} & \lstick{\mathrm{{c_{1,2}} :  }} & \lstick{/_{_{2}}} \cw & \cw & \cw & \cw & \cw & \cw & \cw & \cw & \cw & \cw & \cw & \cw & \cw & \cw & \cw & \cw & \cw & \cw & \cw & \cw & \cw & \cw & \cw & \cw & \dstick{_{_{\hspace{0.0em}{c_1}}}} \cw \ar @{<=} [-12,0] & \dstick{_{_{\hspace{0.0em}{c_2}}}} \cw \ar @{<=} [-1,0] & \cw & \cw\\
 }
  \hspace{5mm}\raisebox{-35mm}{$\Rightarrow$}\hspace{0mm}
\Qcircuit @C=0.2em @R=0.2em @!R { 
\nghost{\mathrm{c_p :  }} & \lstick{\mathrm{c_p :  }} &  \controlo \cw \ar @{-} [1,0]  & \cw & \cw & \cw & \cw & \cw & \cw & \cw & \cw & \cw & \cw & \cw & \cw & \cw & \cw & \cw & \cw & \cw & \cw & \cw & \cw  \\
	 	\nghost{{a}_{1} :  } & \lstick{{a}_{1} :  } & \gate{\mathrm{H}} & \ctrl{1} & \qw & \targ & \gate{\mathrm{\theta}} & \qw & \ctrl{1} & \qw & \qw & \qw & \qw & \qw & \qw & \qw & \qw & \qw & \qw & \qw & \qw & \qw & \qw\\
	 	\nghost{{q}_{1} :  } & \lstick{{q}_{1} :  } & \gate{\mathrm{\theta}} & \targ & \ctrl{1} & \ctrl{-1} & \targ & \gate{\mathrm{\theta}} & \targ & \ctrl{1} & \qw & \qw & \qw & \qw & \qw & \qw & \qw & \qw & \qw & \qw & \qw & \qw & \qw\\
	 	\nghost{{q}_{2} :  } & \lstick{{q}_{2} :  } & \gate{\mathrm{\theta}} & \qw & \targ & \ctrl{1} & \ctrl{-1} & \targ & \gate{\mathrm{\theta}} & \targ & \ctrl{1} & \qw & \qw & \qw & \qw & \qw & \qw & \qw & \qw & \qw & \qw & \qw & \qw\\
	 	\nghost{{q}_{3} :  } & \lstick{{q}_{3} :  } & \gate{\mathrm{\theta}} & \qw & \qw & \targ & \ctrl{1} & \ctrl{-1} & \targ & \gate{\mathrm{\theta}} & \targ & \ctrl{1} & \qw & \qw & \qw & \qw & \qw & \qw & \qw & \qw & \qw & \qw & \qw\\
	 	\nghost{{q}_{4} :  } & \lstick{{q}_{4} :  } & \gate{\mathrm{\theta}} & \qw & \qw & \qw & \targ & \ctrl{1} & \ctrl{-1} & \targ & \gate{\mathrm{\theta}} & \targ & \ctrl{1} & \qw & \qw & \qw & \qw & \qw & \qw & \qw & \qw & \qw & \qw\\
	 	\nghost{{q}_{5} :  } & \lstick{{q}_{5} :  } & \gate{\mathrm{\theta}} & \qw & \qw & \qw & \qw & \targ & \ctrl{1} & \ctrl{-1} & \targ & \gate{\mathrm{\theta}} & \targ & \ctrl{1} & \qw & \qw & \qw & \qw & \qw & \qw & \qw & \qw & \qw\\
	 	\nghost{{q}_{6} :  } & \lstick{{q}_{6} :  } & \gate{\mathrm{\theta}} & \qw & \qw & \qw & \qw & \qw & \targ & \ctrl{1} & \ctrl{-1} & \targ & \gate{\mathrm{\theta}} & \targ & \ctrl{1} & \qw & \qw & \qw & \qw & \qw & \qw & \qw & \qw\\
	 	\nghost{{q}_{7} :  } & \lstick{{q}_{7} :  } & \gate{\mathrm{\theta}} & \qw & \qw & \qw & \qw & \qw & \qw & \targ & \ctrl{1} & \ctrl{-1} & \targ & \gate{\mathrm{\theta}} & \targ & \ctrl{1} & \qw & \qw & \qw & \qw & \qw & \qw & \qw\\
	 	\nghost{{q}_{8} :  } & \lstick{{q}_{8} :  } & \gate{\mathrm{\theta}} & \qw & \qw & \qw & \qw & \qw & \qw & \qw & \targ & \ctrl{1} & \ctrl{-1} & \targ & \gate{\mathrm{\theta}} & \targ & \ctrl{1} & \qw & \qw & \qw & \qw & \qw & \qw\\
	 	\nghost{{q}_{9} :  } & \lstick{{q}_{9} :  } & \gate{\mathrm{\theta}} & \qw & \qw & \qw & \qw & \qw & \qw & \qw & \qw & \targ & \ctrl{1} & \ctrl{-1} & \targ & \gate{\mathrm{\theta}} & \targ & \targ & \ctrl{1} & \qw & \qw & \qw & \qw\\
	 	\nghost{{q}_{10} :  } & \lstick{{q}_{10} :  } & \gate{\mathrm{\theta}} & \qw & \qw & \qw & \qw & \qw & \qw & \qw & \qw & \qw & \targ & \qw & \ctrl{-1} & \qw & \qw & \ctrl{-1} & \targ & \ctrl{1} & \qw & \qw & \qw\\
	 	\nghost{{a}_{2} :  } & \lstick{{a}_{2} :  } & \qw & \qw & \qw & \qw & \qw & \qw & \qw & \qw & \qw & \qw & \qw & \qw & \qw & \qw & \qw & \qw & \qw & \targ & \meter & \qw & \qw\\
	 	\nghost{\mathrm{{c_{2}} :  }} & \lstick{\mathrm{{c_{2}} :  }} & \lstick{/_{_{1}}} \cw & \cw & \cw & \cw & \cw & \cw & \cw & \cw & \cw & \cw & \cw & \cw & \cw & \cw & \cw & \cw & \cw & \cw & \dstick{_{_{\hspace{0.0em}{c_2}}}} \cw \ar @{<=} [-1,0] & \cw & \cw\\
 }}
 \qcref{final_large_depth_parity}
\]
Finally, the qiskit transpiler is used in order to adjust the circuit to the mapping and basic gates of IBM's quantum computer.

\section{Quantum Circuits Optimization: M-method} \label{app:mid_circuit}

The left-hand side of \qc{mid_cir_cir_first_dec} is equivalent to the structure in Fig.~\ref{fig:circuit_1}.(b), with $N=8$, and with a different qubit ordering such that the ancilla qubits are placed in the middle. We wish to find an efficient LNN implementation of this circuit. The right-hand side of \qc{mid_cir_cir_first_dec} is achieved simply by repeatedly applying \qc{identity_cir} to replace sequential long-range CNOT gates with a V-shaped chain of LNN CNOTs.
\[
\scalebox{0.55}{
\Qcircuit @C=0.2em @R=0.2em @!R { 
\nghost{\mathrm{c_p :  }} & \lstick{\mathrm{c_p :  }} & \cw& \cw& \cw& \cw & \cw & \cw& \cw& \cw& \cw &  \control \cw \ar @{-} [5,0] & \cw  & \cw & \cw & \cw & \cw & \cw & \cw & \cw & \cw & \cw & \cw & \cw & \cw & \cw& \cw& \cw& \cw& \cw& \cw & \cw & \cw\\
	 	\nghost{{q}_{1} :  } & \lstick{{q}_{1} :  } & \gate{\mathrm{\theta}} & \ctrl{4} & \qw & \qw & \qw & \qw & \qw & \qw & \qw & \qw & \qw & \qw & \qw & \qw & \qw & \qw & \qw & \ctrl{4} & \qw & \gate{\mathrm{\theta}} & \ctrl{5} & \qw & \qw & \qw & \qw & \qw & \qw & \qw & \qw & \qw & \qw\\
	 	\nghost{{q}_{2} :  } & \lstick{{q}_{2} :  } & \gate{\mathrm{\theta}} & \qw & \ctrl{3} & \qw & \qw & \qw & \qw & \qw & \qw & \qw & \qw & \qw & \qw & \qw & \qw & \qw & \ctrl{3} & \qw & \qw & \gate{\mathrm{\theta}} & \qw & \ctrl{4} & \qw & \qw & \qw & \qw & \qw & \qw & \qw & \qw & \qw\\
	 	\nghost{{q}_{3} :  } & \lstick{{q}_{3} :  } & \gate{\mathrm{\theta}} & \qw & \qw & \ctrl{2} & \qw & \qw & \qw & \qw & \qw & \qw & \qw & \qw & \qw & \qw & \qw & \ctrl{2} & \qw & \qw & \qw & \gate{\mathrm{\theta}} & \qw & \qw & \ctrl{3} & \qw & \qw & \qw & \qw & \qw & \qw & \qw & \qw\\
	 	\nghost{{q}_{4} :  } & \lstick{{q}_{4} :  } & \gate{\mathrm{\theta}} & \qw & \qw & \qw & \ctrl{1} & \qw & \qw & \qw & \qw & \qw & \qw & \qw & \qw & \qw & \ctrl{1} & \qw & \qw & \qw & \qw & \gate{\mathrm{\theta}} & \qw & \qw & \qw & \ctrl{2} & \qw & \qw & \qw & \qw & \qw & \qw & \qw\\
	 	\nghost{{a}_{1} :  } & \lstick{{a}_{1} :  } & \qw & \targ & \targ & \targ & \targ & \targ & \targ & \targ & \targ & \meter & \targ & \targ & \targ & \targ & \targ & \targ & \targ & \targ & \qw & \qw & \qw & \qw & \qw & \qw & \qw & \qw & \qw & \qw & \qw & \qw & \qw\\
	 	\nghost{{a}_{2} :  } & \lstick{{a}_{2} :  } & \qw & \qw & \qw & \qw & \qw & \qw & \qw & \qw & \qw & \qw & \qw & \qw & \qw & \qw & \qw & \qw & \qw & \qw & \qw & \qw & \targ & \targ & \targ & \targ & \targ & \targ & \targ & \targ & \meter & \qw & \qw\\
	 	\nghost{{q}_{5} :  } & \lstick{{q}_{5} :  } & \gate{\mathrm{\theta}} & \qw & \qw & \qw & \qw & \ctrl{-2} & \qw & \qw & \qw  & \qw  & \qw & \qw & \qw & \ctrl{-2} & \qw & \qw & \qw & \qw & \qw & \gate{\mathrm{\theta}} & \qw & \qw & \qw & \qw & \ctrl{-1} & \qw & \qw & \qw  & \qw & \qw & \qw\\
	 	\nghost{{q}_{6} :  } & \lstick{{q}_{6} :  } & \gate{\mathrm{\theta}} & \qw & \qw & \qw & \qw & \qw & \ctrl{-3} & \qw  & \qw & \qw & \qw  & \qw & \ctrl{-3} & \qw & \qw & \qw & \qw & \qw & \qw & \gate{\mathrm{\theta}} & \qw & \qw & \qw & \qw & \qw & \ctrl{-2} & \qw  & \qw & \qw & \qw & \qw\\
	 	\nghost{{q}_{7} :  } & \lstick{{q}_{7} :  } & \gate{\mathrm{\theta}} & \qw & \qw & \qw & \qw & \qw  & \qw & \ctrl{-4} & \qw & \qw & \qw & \ctrl{-4}  & \qw & \qw & \qw & \qw & \qw & \qw & \qw & \gate{\mathrm{\theta}} & \qw & \qw & \qw & \qw & \qw  & \qw & \ctrl{-3} & \qw & \qw & \qw & \qw\\
	 	\nghost{{q}_{8} :  } & \lstick{{q}_{8} :  } & \gate{\mathrm{\theta}} & \qw & \qw & \qw & \qw  & \qw & \qw & \qw & \ctrl{-5} & \qw & \ctrl{-5} & \qw & \qw & \qw  & \qw & \qw & \qw & \qw & \qw & \gate{\mathrm{\theta}} & \qw & \qw & \qw & \qw  & \qw & \qw & \qw & \ctrl{-4} & \qw & \qw & \qw\\
	 	\nghost{\mathrm{{c_{1,2}} :  }} & \lstick{\mathrm{{c_{1,2}} :  }} & \lstick{/_{_{2}}} \cw & \cw & \cw & \cw & \cw & \cw & \cw & \cw & \cw & \dstick{_{_{\hspace{0.0em}{c_1}}}} \cw \ar @{<=} [-6,0] & \cw & \cw & \cw & \cw & \cw & \cw & \cw & \cw & \cw & \cw & \cw & \cw & \cw & \cw & \cw & \cw & \cw & \cw & \dstick{_{_{\hspace{0.0em}{c_2}}}} \cw \ar @{<=} [-5,0] & \cw & \cw\\
 }
   \hspace{5mm}\raisebox{-30mm}{=}\hspace{0mm}
\Qcircuit @C=0.2em @R=0.2em @!R {
\nghost{\mathrm{c_p :  }} & \lstick{\mathrm{c_p :  }} & \cw & \cw & \cw& \cw& \cw& \cw & \cw & \cw& \cw& \cw& \cw &  \control \cw \ar @{-} [5,0] & \cw  & \cw & \cw & \cw & \cw & \cw & \cw & \cw & \cw & \cw & \cw & \cw & \cw & \cw& \cw& \cw& \cw& \cw& \cw & \cw & \cw & \cw & \cw\\
	 	\nghost{{q}_{1} :  } & \lstick{{q}_{1} :  } & \gate{\mathrm{\theta}} & \ctrl{1} & \qw & \qw & \qw & \qw \barrier[0em]{9} & \qw & \qw & \qw & \ctrl{1} & \qw & \qw & \qw & \ctrl{1} & \qw & \qw \barrier[0em]{9} & \qw & \qw & \qw & \qw & \qw & \ctrl{1} & \qw & \gate{\mathrm{\theta}} & \ctrl{1} & \qw & \qw & \qw & \qw \barrier[0em]{9} & \qw & \qw & \qw & \ctrl{1} & \qw & \qw\\
	 	\nghost{{q}_{2} :  } & \lstick{{q}_{2} :  } & \gate{\mathrm{\theta}} & \targ & \ctrl{1} & \qw & \qw & \qw & \qw & \qw & \ctrl{1} & \targ & \qw & \qw & \qw & \targ & \ctrl{1} & \qw & \qw & \qw & \qw & \qw & \ctrl{1} & \targ & \qw & \gate{\mathrm{\theta}} & \targ & \ctrl{1} & \qw & \qw & \qw & \qw & \qw & \ctrl{1} & \targ & \qw & \qw\\
	 	\nghost{{q}_{3} :  } & \lstick{{q}_{3} :  } & \gate{\mathrm{\theta}} & \qw & \targ & \ctrl{1} & \qw & \qw & \qw & \ctrl{1} & \targ & \qw & \qw & \qw & \qw & \qw & \targ & \ctrl{1} & \qw & \qw & \qw & \ctrl{1} & \targ & \qw & \qw & \gate{\mathrm{\theta}} & \qw & \targ & \ctrl{1} & \qw & \qw & \qw & \ctrl{1} & \targ & \qw & \qw & \qw\\
	 	\nghost{{q}_{4} :  } & \lstick{{q}_{4} :  } & \gate{\mathrm{\theta}} & \qw & \qw & \targ & \ctrl{1} & \qw & \qw & \targ & \qw & \qw & \qw & \qw & \qw & \qw & \qw & \targ & \qw & \qw & \ctrl{1} & \targ & \qw & \qw & \qw & \gate{\mathrm{\theta}} & \qw & \qw & \targ & \ctrl{2} & \qw & \qw & \targ & \qw & \qw & \qw & \qw\\
	 	\nghost{{a}_{1} :  } & \lstick{{a}_{1} :  } & \qw & \qw & \qw & \qw & \targ & \targ & \qw & \qw & \qw & \qw & \qw & \meter & \qw & \qw & \qw & \qw & \qw & \targ & \targ & \qw & \qw & \qw & \qw & \qw & \qw & \qw & \qw & \qw & \qw & \qw & \qw & \qw & \qw & \qw & \qw\\
	 	\nghost{{a}_{2} :  } & \lstick{{a}_{2} :  } & \qw & \qw & \qw & \qw & \qw & \qw & \qw & \qw & \qw & \qw & \qw & \qw & \qw & \qw & \qw & \qw & \qw & \qw & \qw & \qw & \qw & \qw & \qw & \qw & \qw & \qw & \targ & \targ & \meter & \qw & \qw & \qw & \qw & \qw & \qw\\
	 	\nghost{{q}_{5} :  } & \lstick{{q}_{5} :  } & \gate{\mathrm{\theta}} & \qw & \qw & \targ & \qw & \ctrl{-2} & \qw & \targ & \qw & \qw & \qw & \qw & \qw & \qw & \qw & \targ & \qw & \ctrl{-2} & \targ & \qw & \qw & \qw & \gate{\mathrm{\theta}} & \qw & \qw & \targ & \ctrl{-1} & \qw & \qw & \qw & \targ & \qw & \qw & \qw & \qw\\
	 	\nghost{{q}_{6} :  } & \lstick{{q}_{6} :  } & \gate{\mathrm{\theta}} & \qw & \targ & \ctrl{-1} & \qw & \qw & \qw & \ctrl{-1} & \targ & \qw & \qw & \qw & \qw & \qw & \targ & \ctrl{-1} & \qw & \qw & \ctrl{-1} & \targ & \qw & \qw & \gate{\mathrm{\theta}} & \qw & \targ & \ctrl{-1} & \qw & \qw & \qw & \qw & \ctrl{-1} & \targ & \qw & \qw & \qw\\
	 	\nghost{{q}_{7} :  } & \lstick{{q}_{7} :  } & \gate{\mathrm{\theta}} & \targ & \ctrl{-1} & \qw & \qw & \qw & \qw & \qw & \ctrl{-1} & \targ & \qw & \qw & \qw & \targ & \ctrl{-1} & \qw & \qw & \qw & \qw & \ctrl{-1} & \targ & \qw & \gate{\mathrm{\theta}} & \targ & \ctrl{-1} & \qw & \qw & \qw & \qw & \qw & \qw & \ctrl{-1} & \targ & \qw & \qw\\
	 	\nghost{{q}_{8} :  } & \lstick{{q}_{8} :  } & \gate{\mathrm{\theta}} & \ctrl{-1} & \qw & \qw & \qw & \qw & \qw & \qw & \qw & \ctrl{-1} & \qw & \qw & \qw & \ctrl{-1} & \qw & \qw & \qw & \qw & \qw & \qw & \ctrl{-1} & \qw & \gate{\mathrm{\theta}} & \ctrl{-1} & \qw & \qw & \qw & \qw & \qw & \qw & \qw & \qw & \ctrl{-1} & \qw & \qw\\
	 	\nghost{\mathrm{{c_{1,2}} :  }} & \lstick{\mathrm{{c_{1,2}} :  }} & \lstick{/_{_{2}}} \cw & \cw & \cw & \cw & \cw & \cw & \cw & \cw & \cw & \cw & \cw & \dstick{_{_{\hspace{0.0em}{c_1}}}} \cw \ar @{<=} [-6,0] & \cw & \cw & \cw & \cw & \cw & \cw & \cw & \cw & \cw & \cw & \cw & \cw & \cw & \cw & \cw & \cw & \dstick{_{_{\hspace{0.0em}{c_2}}}} \cw \ar @{<=} [-5,0] & \cw & \cw & \cw & \cw & \cw & \cw\\
 }}
 \qcref{mid_cir_cir_first_dec}
\]
Finally, all CNOT gates between the first two barriers simply cancel out, and the ones applied after the final measurement can be removed. \qc{mid_cir_cir} can therefore be used instead of \qc{mid_cir_cir_first_dec}.
\[
\scalebox{0.55}{
\Qcircuit @C=0.2em @R=0.2em @!R { 
\nghost{\mathrm{c_p :  }} & \lstick{\mathrm{c_p :  }} & \cw & \cw & \cw& \cw& \cw& \cw &  \control \cw \ar @{-} [5,0] & \cw  & \cw & \cw & \cw & \cw & \cw & \cw & \cw & \cw & \cw & \cw & \cw & \cw\\
	 	\nghost{{q}_{1} :  } & \lstick{{q}_{1} :  } & \gate{\mathrm{\theta}} & \ctrl{1} & \qw & \qw & \qw & \qw & \qw & \qw & \qw & \qw & \qw & \ctrl{1} & \gate{\mathrm{\theta}} & \ctrl{1} & \qw & \qw & \qw & \qw & \qw & \qw\\
	 	\nghost{{q}_{2} :  } & \lstick{{q}_{2} :  } & \gate{\mathrm{\theta}} & \targ & \ctrl{1} & \qw & \qw & \qw & \qw & \qw & \qw & \qw & \ctrl{1} & \targ & \gate{\mathrm{\theta}} & \targ & \ctrl{1} & \qw & \qw & \qw & \qw & \qw\\
	 	\nghost{{q}_{3} :  } & \lstick{{q}_{3} :  } & \gate{\mathrm{\theta}} & \qw & \targ & \ctrl{1} & \qw & \qw & \qw & \qw & \qw & \ctrl{1} & \targ & \qw & \gate{\mathrm{\theta}}  & \qw & \targ & \ctrl{1} & \qw & \qw & \qw & \qw\\
	 	\nghost{{q}_{4} :  } & \lstick{{q}_{4} :  } & \gate{\mathrm{\theta}} & \qw & \qw & \targ & \ctrl{1} & \qw & \qw & \qw & \ctrl{1} & \targ & \qw & \qw & \gate{\mathrm{\theta}}  & \qw & \qw & \targ & \ctrl{2} & \qw & \qw & \qw\\
	 	\nghost{{a}_{1} :  } & \lstick{{a}_{1} :  } & \qw & \qw & \qw & \qw & \targ & \targ & \meter & \targ & \targ & \qw & \qw & \qw & \qw & \qw & \qw & \qw & \qw & \qw & \qw & \qw\\
	 	\nghost{{a}_{2} :  } & \lstick{{a}_{2} :  } & \qw & \qw & \qw & \qw & \qw & \qw & \qw & \qw & \qw & \qw & \qw & \qw & \qw & \qw & \qw & \targ & \targ & \meter & \qw & \qw\\
	 	\nghost{{q}_{5} :  } & \lstick{{q}_{5} :  } & \gate{\mathrm{\theta}} & \qw & \qw & \targ & \qw & \ctrl{-2} & \qw & \ctrl{-2} & \targ & \qw & \qw  & \gate{\mathrm{\theta}} & \qw & \qw & \targ & \ctrl{-1} & \qw & \qw & \qw & \qw\\
	 	\nghost{{q}_{6} :  } & \lstick{{q}_{6} :  } & \gate{\mathrm{\theta}} & \qw & \targ & \ctrl{-1} & \qw & \qw & \qw & \qw & \ctrl{-1} & \targ & \qw & \gate{\mathrm{\theta}}  & \qw & \targ & \ctrl{-1} & \qw & \qw & \qw & \qw & \qw\\
	 	\nghost{{q}_{7} :  } & \lstick{{q}_{7} :  } & \gate{\mathrm{\theta}} & \targ & \ctrl{-1} & \qw & \qw & \qw & \qw & \qw & \qw & \ctrl{-1} & \targ & \gate{\mathrm{\theta}} & \targ & \ctrl{-1} & \qw & \qw & \qw & \qw & \qw & \qw\\
	 	\nghost{{q}_{8} :  } & \lstick{{q}_{8} :  } & \gate{\mathrm{\theta}} & \ctrl{-1} & \qw & \qw & \qw & \qw & \qw & \qw & \qw & \qw & \ctrl{-1} & \gate{\mathrm{\theta}} & \ctrl{-1} & \qw & \qw & \qw & \qw & \qw & \qw & \qw\\
	 	\nghost{\mathrm{{c_{1,2}} :  }} & \lstick{\mathrm{{c_{1,2}} :  }} & \lstick{/_{_{2}}} \cw & \cw & \cw & \cw & \cw & \cw & \dstick{_{_{\hspace{0.0em}{c_1}}}} \cw \ar @{<=} [-6,0] & \cw & \cw & \cw & \cw & \cw & \cw & \cw & \cw & \cw & \cw & \dstick{_{_{\hspace{0.0em}{c_2}}}} \cw \ar @{<=} [-5,0] & \cw & \cw\\
 }}
 \qcref{mid_cir_cir}
\]

There are three remaining next-nearest neighbor CNOT gates which skip the ancillas. The LNN decomposition of these is trivial.
This circuit over $N$ qubits is applied using $3N+O(1)$ LNN CNOT gates in depth $1.5N+O(1)$. 

As mentioned in Sec.~\ref{sec:implementation}, in case the first measurement is not applied ($c_p=0$), we will apply a measurement on an uncorrelated qubit, and only after this measurement has been completed, the following gate instructions will apply. This guarantees that in both cases, the circuit runtime will be the same, which in turn reduces CD.


\section{Classical Disturbance}
\label{app:Classical_Disturbance}

In order to detect CDs in the QCs using the methods presented in Sec.~\ref{sec:implementation}, one can proceed by either canceling the entire entangling parity dynamics ($\hat{U}_p$), which is then mitigated via the H-method, or removing the mid-circuit measurement, which is then mitigated in the M-method by measuring another uncorrelated qubit. 

For the former, the circuits (achieved similarly to Appendix~\ref{app:quantum_circ}) differ significantly between the two sub protocols (single and double measurement): 

\[
\scalebox{0.55}{
\Qcircuit @C=0.2em @R=0.2em @!R{ 
	 	\nghost{{a}_{1} :  } & \lstick{{a}_{1} :  } & \qw & \ctrl{1} & \qw & \targ & \gate{\mathrm{\theta}} & \qw & \ctrl{1} & \qw & \qw & \qw & \qw & \qw & \qw & \qw & \qw & \qw & \qw & \qw & \qw & \qw & \qw\\
	 	\nghost{{q}_{1} :  } & \lstick{{q}_{1} :  } & \gate{\mathrm{\theta}} & \targ & \ctrl{1} & \ctrl{-1} & \targ & \gate{\mathrm{\theta}} & \targ & \ctrl{1} & \qw & \qw & \qw & \qw & \qw & \qw & \qw & \qw & \qw & \qw & \qw & \qw & \qw\\
	 	\nghost{{q}_{2} :  } & \lstick{{q}_{2} :  } & \gate{\mathrm{\theta}} & \qw & \targ & \ctrl{1} & \ctrl{-1} & \targ & \gate{\mathrm{\theta}} & \targ & \ctrl{1} & \qw & \qw & \qw & \qw & \qw & \qw & \qw & \qw & \qw & \qw & \qw & \qw\\
	 	\nghost{{q}_{3} :  } & \lstick{{q}_{3} :  } & \gate{\mathrm{\theta}} & \qw & \qw & \targ & \ctrl{1} & \ctrl{-1} & \targ & \gate{\mathrm{\theta}} & \targ & \ctrl{1} & \qw & \qw & \qw & \qw & \qw & \qw & \qw & \qw & \qw & \qw & \qw\\
	 	\nghost{{q}_{4} :  } & \lstick{{q}_{4} :  } & \gate{\mathrm{\theta}} & \qw & \qw & \qw & \targ & \ctrl{1} & \ctrl{-1} & \targ & \gate{\mathrm{\theta}} & \targ & \ctrl{1} & \qw & \qw & \qw & \qw & \qw & \qw & \qw & \qw & \qw & \qw\\
	 	\nghost{{q}_{5} :  } & \lstick{{q}_{5} :  } & \gate{\mathrm{\theta}} & \qw & \qw & \qw & \qw & \targ & \ctrl{1} & \ctrl{-1} & \targ & \gate{\mathrm{\theta}} & \targ & \ctrl{1} & \qw & \qw & \qw & \qw & \qw & \qw & \qw & \qw & \qw\\
	 	\nghost{{q}_{6} :  } & \lstick{{q}_{6} :  } & \gate{\mathrm{\theta}} & \qw & \qw & \qw & \qw & \qw & \targ & \ctrl{1} & \ctrl{-1} & \targ & \gate{\mathrm{\theta}} & \targ & \ctrl{1} & \qw & \qw & \qw & \qw & \qw & \qw & \qw & \qw\\
	 	\nghost{{q}_{7} :  } & \lstick{{q}_{7} :  } & \gate{\mathrm{\theta}} & \qw & \qw & \qw & \qw & \qw & \qw & \targ & \ctrl{1} & \ctrl{-1} & \targ & \gate{\mathrm{\theta}} & \targ & \ctrl{1} & \qw & \qw & \qw & \qw & \qw & \qw & \qw\\
	 	\nghost{{q}_{8} :  } & \lstick{{q}_{8} :  } & \gate{\mathrm{\theta}} & \qw & \qw & \qw & \qw & \qw & \qw & \qw & \targ & \ctrl{1} & \ctrl{-1} & \targ & \gate{\mathrm{\theta}} & \targ & \ctrl{1} & \qw & \qw & \qw & \qw & \qw & \qw\\
	 	\nghost{{q}_{9} :  } & \lstick{{q}_{9} :  } & \gate{\mathrm{\theta}} & \qw & \qw & \qw & \qw & \qw & \qw & \qw & \qw & \targ & \ctrl{1} & \ctrl{-1} & \targ & \gate{\mathrm{\theta}} & \targ & \targ & \ctrl{1} & \qw & \qw & \qw & \qw\\
	 	\nghost{{q}_{10} :  } & \lstick{{q}_{10} :  } & \gate{\mathrm{\theta}} & \qw & \qw & \qw & \qw & \qw & \qw & \qw & \qw & \qw & \targ & \qw & \ctrl{-1} & \qw & \qw & \ctrl{-1} & \targ & \ctrl{1} & \qw & \qw & \qw\\
	 	\nghost{{a}_{2} :  } & \lstick{{a}_{2} :  } & \qw & \qw & \qw & \qw & \qw & \qw & \qw & \qw & \qw & \qw & \qw & \qw & \qw & \qw & \qw & \qw & \qw & \targ & \meter & \qw & \qw\\
	 	\nghost{\mathrm{{c_{2}} :  }} & \lstick{\mathrm{{c_{2}} :  }} & \lstick{/_{_{1}}} \cw & \cw & \cw & \cw & \cw & \cw & \cw & \cw & \cw & \cw & \cw & \cw & \cw & \cw & \cw & \cw & \cw & \cw & \dstick{_{_{\hspace{0.0em}{c_2}}}} \cw \ar @{<=} [-1,0] & \cw & \cw\\
 }}
 \hspace{10mm}\raisebox{-37mm}{;}\hspace{10mm}
 \scalebox{0.55}{
\Qcircuit @C=0.2em @R=0.2em @!R {
\nghost{{a}_{1} :  } & \lstick{{a}_{1} :  } & \qw & \qw & \qw & \qw & \qw & \qw & \qw & \qw & \qw & \qw & \qw & \qw & \qw & \qw\\
	 	\nghost{{q}_{1} :  } & \lstick{{q}_{1} :  } & \gate{\mathrm{2\theta}} & \ctrl{1} & \qw & \qw & \qw & \qw & \qw & \qw & \qw & \qw & \qw & \qw & \qw & \qw\\
	 	\nghost{{q}_{2} :  } & \lstick{{q}_{2} :  } & \gate{\mathrm{2\theta}} & \targ & \ctrl{1} & \qw & \qw & \qw & \qw & \qw & \qw & \qw & \qw & \qw & \qw & \qw\\
	 	\nghost{{q}_{3} :  } & \lstick{{q}_{3} :  } & \gate{\mathrm{2\theta}} & \qw & \targ & \ctrl{1} & \qw & \qw & \qw & \qw & \qw & \qw & \qw & \qw & \qw & \qw\\
	 	\nghost{{q}_{4} :  } & \lstick{{q}_{4} :  } & \gate{\mathrm{2\theta}} & \qw & \qw & \targ & \ctrl{1} & \qw & \qw & \qw & \qw & \qw & \qw & \qw & \qw & \qw\\
	 	\nghost{{q}_{5} :  } & \lstick{{q}_{5} :  } & \gate{\mathrm{2\theta}} & \qw & \qw & \qw & \targ & \ctrl{1} & \qw & \qw & \qw & \qw & \qw & \qw & \qw & \qw\\
	 	\nghost{{q}_{6} :  } & \lstick{{q}_{6} :  } & \gate{\mathrm{2\theta}} & \qw & \qw & \qw & \qw & \targ & \ctrl{1} & \qw & \qw & \qw & \qw & \qw & \qw & \qw\\
	 	\nghost{{q}_{7} :  } & \lstick{{q}_{7} :  } & \gate{\mathrm{2\theta}} & \qw & \qw & \qw & \qw & \qw & \targ & \ctrl{1} & \qw & \qw & \qw & \qw & \qw & \qw\\
	 	\nghost{{q}_{8} :  } & \lstick{{q}_{8} :  } & \gate{\mathrm{2\theta}} & \qw & \qw & \qw & \qw & \qw & \qw & \targ & \ctrl{1} & \qw & \qw & \qw & \qw & \qw\\
	 	\nghost{{q}_{9} :  } & \lstick{{q}_{9} :  } & \gate{\mathrm{2\theta}} & \qw & \qw & \qw & \qw & \qw & \qw & \qw & \targ & \ctrl{1} & \qw & \qw & \qw & \qw\\
	 	\nghost{{q}_{10} :  } & \lstick{{q}_{10} :  } & \gate{\mathrm{2\theta}} & \qw & \qw & \qw & \qw & \qw & \qw & \qw & \qw & \targ & \ctrl{1} & \qw & \qw & \qw\\
	 	\nghost{{a}_{2} :  } & \lstick{{a}_{2} :  } & \qw & \qw & \qw & \qw & \qw & \qw & \qw & \qw & \qw & \qw & \targ & \meter & \qw & \qw\\
	 	\nghost{\mathrm{{c_2} :  }} & \lstick{\mathrm{{c_2} :  }} & \lstick{/_{_{1}}} \cw & \cw & \cw & \cw & \cw & \cw & \cw & \cw & \cw & \cw & \cw & \dstick{_{_{\hspace{0.0em}c_2}}} \cw \ar @{<=} [-1,0] & \cw & \cw\\
 }}
  \qcref{old_cir_cir}
\]
which introduces a large CDs, as reported in Fig.~\ref{fig:IBM_results}.(c).

\end{document}